\newcommand{\diracslash}[1]{#1\llap{/\kern2pt}}

\newcommand{\be}{\begin{equation}}
\newcommand{\ee}{\end{equation}}
\newcommand{\bea}{\begin{eqnarray}}
\newcommand{\eea}{\end{eqnarray}}
\newcommand{\ba}[1]{\begin{array}{#1}}
\newcommand{\ea}{\end{array}}

\newcommand{\bt}{\begin{tabular}}
\newcommand{\et}{\end{tabular}}

\newcommand{\beas}{\begin{eqnarray*}}
\newcommand{\eeas}{\end{eqnarray*}}

\documentclass[preprint,prd,aps,floats,nofootinbib,floatfix]{revtex4}

\DeclareSymbolFont{rsfs}{U}{rsfs}{m}{n}
\DeclareSymbolFontAlphabet{\mathrsfs}{rsfs}

\usepackage{graphicx}
\begin{document}

\title{Heavy Scalar, Vector and Axial-Vector Mesons 
in Hot and Dense Nuclear Medium}
\author{Arvind Kumar}
\email{iitd.arvind@gmail.com, kumara@nitj.ac.in}
\affiliation{Department of Physics, Dr. B R Ambedkar National Institute of Technology Jalandhar, 
 Jalandhar -- 144011,Punjab, India}

\def\be{\begin{equation}}
\def\ee{\end{equation}}
\def\bearr{\begin{eqnarray}}
\def\eearr{\end{eqnarray}}
\def\zbf#1{{\bf {#1}}}
\def\bfm#1{\mbox{\boldmath $#1$}}
\def\hf{\frac{1}{2}}
\def\kp{\zbf k+\frac{\zbf q}{2}}
\def\km{-\zbf k+\frac{\zbf q}{2}}
\def\hwo{\hat\omega_1}
\def\hwt{\hat\omega_2}

\begin{abstract}

In this work we shall investigate the mass modifications of scalar mesons $\left( D_{0}, B_{0}\right)$,
 vector mesons $\left( D^{\ast}, B^{\ast}\right)$ 
and axial-vector mesons $\left( D_{1}, B_{1}\right)$ at finite density and temperature of the nuclear medium.
The above mesons are modified in the nuclear medium through the
 modification of quark and gluon condensates.
We shall find the medium modification of  quark  and gluon condensates within chiral SU(3) model
 through the medium modification of scalar-isoscalar fields
$\sigma$ and $\zeta$  at finite density and temperature.
These medium modified quark and gluon condensates will further be
used
through QCD sum rules for the evaluation of 
in-medium properties of above mentioned scalar, vector
and axial vector mesons.
We shall also discuss the effects of density and temperature of the
nuclear medium on the scattering lengths of above scalar, vector and
axial-vector mesons. The study of the medium modifications of 
above mesons may be helpful for understanding their production
rates in heavy-ion collision experiments. 
The results of present investigations of medium modifications of
scalar, vector and axial-vector mesons at
finite density and temperature can be verified in the 
Compressed Baryonic Matter (CBM) experiment of FAIR facility at
GSI, Germany. 


\end{abstract}

\maketitle

\section{Introduction}
The motive behind the heavy-ion collision experiments
at different experimental facilities is
to explore the different phases of QCD phase diagram.
These experiments help us to understand the nuclear matter
properties for different values of temperatures and densities.
The hadronic matter produced in 
heavy-ion collisions may undergo different phase transitions
e.g. liquid-gas phase transition, the kaon condensation, the restoration of chiral symmetry
and may be the formation of quark gluon plasma \cite{le1,kap1,kap2,br2}.
The Compressed Baryonic Matter (CBM) experiment of the FAIR project at GSI, Germany
may explore the phase of hadronic matter at high baryon densities and moderate temperatures. 
These kind of phases may exist in the
 compact astrophysical objects e.g. neutron stars.
The property of restoration of chiral symmetry is closely related to the
medium modifications of hadrons \cite{br2}. 
 The medium modifications of 
Kaons, D mesons and light vector mesons had been studied using different
 theoretical approaches e.g. chiral model \cite{paper3,kmeson,isoamss,isoamss2,
 amarind,amdmeson,amarvind,amavstranged}, 
 QCD sum rules \cite{leemorita2,moritalee2008,klingl,koike1,higler1,haya1}
and coupled channel approach
 \cite{oset635,koch337,tolos2}.
Due to interactions the properties of 
hadrons in  the medium are found to be different as compared to 
their free space properties.

 The medium modifications
of heavy scalar, vector and axial-vector mesons 
at finite density and temperature of the medium
had been studied very rarely \cite{wang1,wang2,tolos1,higlersc}.
In the present investigation we shall study the mass modifications
of heavy scalar mesons  $\left( D_{0}, B_{0}\right)$,
 vector mesons
  $\left( D^{\ast}, B^{\ast}\right)$ 
and axial vector mesons $\left( D_{1}, B_{1}\right)$ at finite densities and temperatures.
The study of in-medium properties of scalar, vector and 
axial-vector mesons will be helpful to understand their experimental production rates. 
 The
medium modification of charmed mesons may modify 
the experimental production of 
ground state charmonium $J/\psi$ and the excited
 charmonium states $(\psi^{\prime}$ and $\chi_{c})$.
The charmonium, $J/\psi$ may be produced due to
 the decay of the higher charmonium states. 
However the vacuum threshold value of heavy meson 
pairs lies above the vacuum mass of the excited 
charmonium states. Now if these heavy mesons
 get modified (undergo mass drop in the medium)
then the excited charmoium states may decay to the
 open charmed meson pairs instead of decaying to 
the ground state charmonium.  Thus to understand 
the production of charmonium states in heavy-ion collisions
it is 
very necessary to study the medium 
modification of the heavy scalar, vector 
and axial-vector mesons. The medium modifications of heavy vector mesons may also help us in
understanding the dilepton spectra produced in heavy-ion collision experiments \cite{vec1,vec2,vec3}. The dileptons are
considered as interesting probe to study the evolution of matter produced
in heavy ion collision experiments as they do not undergo strong 
interactions in the medium. In ref. \cite{andr1} the production
of open charm and charmonium  in hot hadronic medium had
been investigated using the statistical hadronization model at
SPS/FAIR energies.
In this work it was observed that the medium modifications of
charmed hadrons do not lead to appreciable changes in cross-section
for D mesons production. This is because of large 
charm quark mass and different times scales
for charm quark and charm hadron production. However, the charmonia
yield is effected appreciably due to in-medium modifications.

 The properties of scalar charm resonances $D_{s0}(2317)$ and $D_0(2400)$
 and hidden charm resonance, $X(3700)$ had been studied in 
 ref. \cite{tolos1} using coupled channel approach. In these studies the $D_{s0}(2317)$ and
 $X(3700)$ were found to undergo a width of about 100 and 200 MeV respectively
 at nuclear matter density. However, for the $D_0(2400)$ mesons there was
 already large width of resonance in the free space and 
 the medium effect were found to be weak as compared to 
 $D_{s0}(2317)$  and $X(3700)$.
 In ref.\cite{higler1} the mass splitting  of $D$-$\bar{D}$ and $B$-$\bar{B}$ mesons 
 had been studied using the 
 QCD sum rules in the cold nuclear matter and the 
 calculated values of mass splitting at nuclear saturation
  density were 60 and 130 MeV respectively.
  The Borel transformed QCD sum rules had also been used to study the properties of pseudoscalar
  $D$ mesons \cite{haya1} and vector mesons, $\rho$, $\omega$ and $\phi$ \cite{koike1}.
The properties of the scalar mesons  $\left( D_{0}, B_{0}\right)$
 in the cold nuclear matter using QCD sum rules have
been investigated in ref. \cite{wang1}. The vectors mesons $\left( D^{*}, B^{*}\right)$  and axial
 vector mesons $\left( D_{1}, B_{1}\right)$ had also been studied using QCD sum rules in cold nuclear matter
 in ref. \cite{wang2}. Note that in ref. \cite{wang1} and \cite{wang2} the  properties of the
 meson were investigated at zero temperature and at normal
nuclear matter density. However in the present investigation
 we shall find the in-medium masses of the 
scalar $\left( D_{0}, B_{0}\right)$ and vector $\left( D^{*}, B^{*}\right)$
and axial vector $\left( D_{1}, B_{1}\right)$ mesons at finite temperatures as well as at
 the densities greater than the nuclear saturation density.

In the present work to investigate the properties of scalar, vector and
axial-vector mesons we shall use the QCD sum rules and chiral SU(3) model \cite{paper3}.
Within QCD sum rules, the in-medium properties of  mesons are related to the
in-medium properties of quark and gluon condensates.
We shall investigate the in-medium properties of quark and gluon condensates 
using the chiral SU(3) model. Using chiral SU(3)
 model we shall find the values  of quark
and gluon condensates at finite values of temperatures and baryonic densities.
These values of condensates will further be used to find the medium modification of mesons using 
QCD sum rules. The chiral SU(3) model along with QCD sum rules
had been used in the literature to 
investigate the in-medium modification of
 the charmonium states $J/\psi$ and $\eta_c$ \cite{charmmass2}.

The present article is organized as follows: In section II we shall 
give a brief review of 
chiral SU(3) model. Then in section III we shall discuss that
how we will evaluate the in-medium modifications of the scalar,
 vector and axial-vector mesons
within QCD sum rules and using the properties of quark and
gluon condensates as evaluated in the chiral SU(3) model.
In section IV we shall discuss the results of the present 
investigation and finally in section V we shall give
a brief summary of present work.
 
 \section{Chiral SU(3) model}
In this section we shall briefly review the chiral SU(3) model 
 used in the present investigation for the
 in-medium properties of heavy mesons. The chiral SU(3) model 
  is based on the broken scale invariance and non-linear realization of chiral symmetry 
  \cite{hartree,kristof1,weinberg,coleman,bardeen}. The
 model involve the Lagrangian densities describing e.g.
 kinetic energy terms, baryon-meson interactions,
 self interactions of scalar mesons, vector mesons, symmetry breaking terms
 and also the scale invariance breaking logarithmic
 potential terms. 

For the investigation of hadron properties at finite temperature and densities
 we use the mean  field approximation.
Under this approximation all the meson fields are treated as classical fields
and only the scalar and the vector fields 
contribute to the baryon-meson interactions.
From the interaction Lagrangian densities, using the mean-field approximation,
  we derive the equations of motions for
the scalar fields  $\sigma$ and $\zeta$ and the dilaton field, $\chi$ in isospin symmetric
nuclear medium. We solve these coupled equations 
to obtain the density and temperature 
dependence of scalar fields  $\sigma$ and $\zeta$ and the dilaton field, $\chi$
in isospin symmetric  nuclear medium \cite{amarvind}.
The concept of broken scale invariance leading to the trace anomaly 
in (massless) QCD, $\theta_{\mu}^{\mu} = \frac{\beta_{QCD}}{2g} 
{G^a}_{\mu\nu} G^{\mu\nu a}$, where $G_{\mu\nu}^{a} $ is the 
gluon field strength tensor of QCD, is simulated in the effective 
Lagrangian at tree level \cite{sche1} through the introduction of 
the scale breaking terms \cite{amarvind}.
Within chiral SU(3) model the scale breaking terms are written in terms
of the dilaton field $\chi$ and also the scalar fields $\sigma$ and
$\zeta$.  From this we obtain the energy momentum tensor
and this is compared with the energy momentum tensor of QCD which is
written in terms of gluon condensates.
In this way we extract the value of gluon condensates in terms 
of the scalar fields $\sigma$ and $\zeta$ and the dilaton field, $\chi$
and is given by the following equation \cite{amarvind},
\begin{eqnarray}
\left\langle  \frac{\alpha_{s}}{\pi} {G^a}_{\mu\nu} {G^a}^{\mu\nu} 
\right\rangle = \frac{8}{9} \Bigg [(1 - d) \chi^{4}
+\left( \frac {\chi}{\chi_{0}}\right)^{2} 
\left( m_{\pi}^{2} f_{\pi} \sigma 
+ \big( \sqrt {2} m_{K}^{2}f_{K} - \frac {1}{\sqrt {2}} 
m_{\pi}^{2} f_{\pi} \big) \zeta \right) \Bigg ],
\label{chiglu}
\end{eqnarray}

where the value of parameter d is 0.064 \cite{amarind}, $m_\pi$ and $m_K$
denote the masses of pions and kaons and have values 139 and 498 MeV respectively.
$f_\pi$ and $f_K$ are the decay constants
having values 93.3 and 122 MeV respectively.
The symbols $\sigma$, $\zeta$ and $\chi$
denote the non-strange scalar-isoscalar field, strange scalar-isoscalar field
and the dilaton field respectively. $\chi_0$ denotes the
value of the dilaton field in vacuum. The vacuum values of 
$\sigma$, $\zeta$ and $\chi$ are -93.3, -106.6 and 409.8 MeV respectively.
Note that in above equation the gluon condensate is written considering
finite quark masses. If we have massless QCD, then only first term written
in terms of dilaton field $\chi$ contributes to
the gluon condensates.
Using above equation we obtain the values of scalar gluon condensates
at different values of densities and 
temperatures of the nuclear medium.

\section{QCD sum rules for scalar $\left( D_{0}, B_{0}\right)$,
 vector  $\left( D^{\ast}, B^{\ast}\right)$ 
and axial vector  $\left( D_{1}, B_{1}\right)$ mesons}

In this section we shall discuss the QCD sum rules \cite{wang1,wang2} which will be
used later along with the chiral SU(3) model for the evaluation of 
in-medium properties of scalar, vector and axial vector mesons. 
To find the mass modification of above discussed heavy mesons we shall
 use the two-point correlation function $\Pi_{\mu\nu}(q)$,
 \begin{eqnarray}
\Pi_{\mu\nu}(q) &=& i\int d^{4}x\ e^{iq \cdot x} \langle T\left\{J_\mu(x)J_\nu^{\dag}(0)\right\} \rangle_{\rho_B, T} \,.
 \end{eqnarray}
 In above equation $J_\mu(x)$ denotes the isospin averaged current, $x = x^\mu = (x^0,\textbf{x})$ is the four coordinate,
  $q = q^\mu = (q^0,\textbf{q})$ is four momentum and $T$ denotes the
 time ordered operation on the product of quantities in the brackets. From above definition it is clear
 that the two point correlation function is actually a Fourier transform 
 of the expectation value of the
 time ordered product of two currents.
 The two-point correlation function for the scalar mesons is defined as,
 \begin{eqnarray}
\Pi (q) &=& i\int d^{4}x\ e^{iq \cdot x} \langle T\left\{J (x)J ^{\dag}(0)\right\} \rangle_{\rho_B, T} \,.
 \end{eqnarray}
 
 For the scalar, vector and axial vector mesons isospin average currents are given by the
 expressions
 \begin{eqnarray}
 J (x) &= &J ^\dag(x) =\frac{\bar{c}(x) q(x)+\bar{q}(x) c(x)}{2}\, , \nonumber\\
 \label{scalarcurrent}
 \end{eqnarray}
 
\begin{eqnarray}
 J_\mu(x) &= &J_\mu^\dag(x) =\frac{\bar{c}(x)\gamma_\mu q(x)+\bar{q}(x)\gamma_\mu c(x)}{2}\, , \nonumber\\
 \label{vectorcurrent}
 \end{eqnarray}
 and
 \begin{eqnarray}
  J_{5\mu}(x) &=&J_{5\mu}^\dag(x) =\frac{\bar{c}(x)\gamma_\mu \gamma_5q(x)+\bar{q}(x)\gamma_\mu\gamma_5 c(x)}{2}\,,
  \label{axialcurrent}
\end{eqnarray}
respectively.
 Note that in above equations $q$ denotes the light $u$ or $d$ quark
  whereas $c$ denotes the heavy charm quark.
Note that in the present work, instead of considering the mass splitting 
between particles and antiparticles, 
we emphasize on the mass shift of iso-doublet $D$ and B mesons
as a whole and therefore we consider the average in the definitions of
scalar, vector and iso-vector currents
which is referred as centroid \citep{haya1}.
To find the mass splitting of particles
and antiparticles in the nuclear medium one has to consider the even and 
odd part of QCD sum rules \cite{higler1}.
For example, in ref. \cite{higler1} the mass splitting between 
pseudoscalar $D$
and $\bar{D}$ mesons was investigated using the even and odd QCD sum rules
whereas in \cite{haya1,wang1,wang2} the mass-shift of $D$ mesons was investigated
under centroid approximation.   
  
 The two point correlation function can be decomposed into
  the vacuum part, a static one-nucleon part and pion bath contribution i.e. we can write
 \begin{eqnarray}
\Pi_{\mu\nu}(q) &=&\Pi^{0}_{\mu\nu}(q)+ \frac{\rho_B}{2M_N}T^{N}_{\mu\nu}(q) +
\Pi^{P.B.}_{\mu\nu}(q)\, ,
\label{pibn}
 \end{eqnarray}
where
\begin{eqnarray}
T^{N}_{\mu\nu}(\omega,\mbox{\boldmath $q$}\,) &=&i\int d^{4}x e^{iq\cdot x}\langle N(p)|
T\left\{J_\mu(x)J_\nu^{\dag}(0)\right\} |N(p) \rangle\,.
\end{eqnarray}
In above equation $|N(p)\rangle$ denotes the isospin and spin averaged static nucleon state
with the four-momentum $p = (M_{N},0)$. The state is normalized as 
 $\langle N(\mbox{\boldmath $p$})|N(\mbox{\boldmath
$p$}')\rangle = (2\pi)^{3} 2p_{0}\delta^{3}(\mbox{\boldmath
$p$}-\mbox{\boldmath $p$}')$.  The third term, $\Pi^{P.B.}_{\mu\nu}(q)$
in equation (\ref{pibn}) gives the contribution from pion bath at finite temperature.
Note that in the present work instead of considering the
contribution of pion bath the effects of finite temperature of the
nuclear matter on the properties of $D$ and $B$ mesons 
will be evaluated through the temperature dependence
of scalar fields $\sigma$, $\zeta$ and $\chi$.  
 The temperature dependence  
of scalar fields $\sigma$, $\zeta$ and $\chi$
modify the nucleon properties in the medium and these modified nucleons
further modify the in-medium properties of $D$ and $B$ mesons at finite temperature
and density. In literature the properties of kaons and antikaons, $D$ mesons
and charmonium had been studied at finite temperature of the nuclear matter
using the above mentioned scalar fields $\sigma$, $\zeta$ and $\chi$ \cite{isoamss,amarvind,amavstranged,charmmass2}.
 
As discussed in Ref. \cite{wang2}, in the limit of the $3$-vector  $\mbox{\boldmath $q$}\rightarrow {\bf 0}$, the
correlation functions $T_{N}(\omega,\mbox{\boldmath $q$}\,)$ can be related to the $D^{*}N$ and $D_{1}N$ 
 scattering T-matrices. Thus we write \cite{wang2}
 \begin{eqnarray}
 {{\cal T}_{D^*N}}(M_{D^*},0)= 8\pi(M_N+M_{D^*})a_{D^*} \nonumber\\
{{\cal T}_{D_1N}}(M_{D_1},0) = 8\pi(M_N+M_{D_1})a_{D_1}
\end{eqnarray}
In above equation $a_{D^*}$ and $a_{D_1}$ are the scattering lengths of $D^*N$ and $D_1N$
respectively.
Similarly, we can also write the scattering $T$ matrix corresponding to $D_{0}N$ ($D_{0}$ is a scalar meson) 
in terms of the scattering lengths \cite{wang1},
 \begin{eqnarray}
 {{\cal T}_{D_{0}N}}(M_{D_{0}},0)= 8\pi(M_N+M_{D_0})a_{D_0}.
\end{eqnarray}
Near the pole positions of the scalar, vector and axial vector mesons
the phenomenological spectral densities can be parameterized with three unknown parameters
$a, b$ and $c$ i.e. we write \cite{wang1,wang2,haya1}
\begin{eqnarray}
\rho(\omega,0) &=& -\frac{f_{D_0/D^*/D_1}^2M_{D_0/D^*/D_1}^2}{\pi}
 \mbox{Im} \left[\frac{{{\cal T}_{D_0/D^*/D_1N}}(\omega,{\bf 0})}{(\omega^{2}-
M_{D_0/D^*/D_1}^2+i\varepsilon)^{2}} \right]\nonumber\\
&+& \cdots = a\,\frac{d}{d\omega^2}\delta^{'}(\omega^{2}-M_{D_0/D^*/D_1}^2)
 +
b\,\delta(\omega^{2}-M_{D_0/D^*/D_1}^2) + c\,\theta(\omega^{2}-s_{0})\,.
\label{a1}
\end{eqnarray}
The term denoted by $...$ represent the continuum contributions.
The first term denotes the double-pole term and corresponds to 
the on-shell effects of the T-matrices,

\begin{eqnarray}
a=-8\pi(M_N+M_{D_0/D^*/D_1})
 a_{D_0/D^*/D_1}f_{D_0/D^*/D_1}^2M_{D_0/D^*/D_1}^2\,.
\label{a2}
\end{eqnarray}

Now we shall write the relation between the scattering length of mesons and
their in-medium mass-shift. For this first we note that
the shift of squared mass of mesons can be written in 
terms of the parameter $a$ appearing in equation (\ref{a1}) through relation \cite{koike1},
\begin{eqnarray}
&&\Delta m_{D_0/D^*/D_1}^{2} = \frac{\rho_N }{2M_N} \frac{a }{f_{D_0/D^*/D_1}^2M_{D_0/D^*/D_1}^2} \nonumber\\
&=& \frac{\rho_N }{2M_N}8\pi(M_N+M_{D_0/D^*/D_1})a_{D_0/D^*/D_1}\,, 
\label{massshiftsq}
\end{eqnarray}
where in the last term we used equation (\ref{a2}).
The mass shift is now defined by the relation
\begin{eqnarray}
\delta M_{D_0/D^*/D_1} = \sqrt{m_{D_0/D^*/D_1}^2+\Delta m_{D_0/D^*/D_1}^{2}}
- m_{D_0/D^*/D_1} .
\label{massshift}
\end{eqnarray}
The second term in equation (\ref{a1}) denotes  the single-pole term,
and corresponds to the off-shell (i.e. $\omega^2\neq M_{D_0/D^*/D_1}^2$) effects of the $T$-matrices. The
third term denotes   the continuum term or the  remaining effects,
where, $s_{0}$, is the continuum threshold.
The continuum threshold parameter $s_0$ define the scale below which
the continuum contribution
vanishes \cite{kwon1}.

It can be observed from  equations
 (\ref{massshiftsq}) and (\ref{massshift}) that if we want to find the value of mass shift of 
mesons then we first need to find the value of unknown parameter $a$.
For this we proceed as follows: we note that in the low energy limit, 
$\omega\rightarrow 0$, the
$T_{N}(\omega,{\bf 0})$ is equivalent to the Born term $T_{D_0/D^*/D_1N}^{\rm
Born}(\omega,{\bf 0})$. We take
into account the Born term at the phenomenological side,
\begin{eqnarray}
T_{N}(\omega^2)&=&T_{D_0/D^*/D_1N}^{\rm
Born}(\omega^2)+\frac{a}{(M_{D_0/D^*/D_1}^2-\omega^2)^2}\nonumber\\
&+&\frac{b}{M_{D_0/D^*/D_1}^2-\omega^2}+\frac{c}{s_0-\omega^2}\, ,
\label{bornconst}
\end{eqnarray}
with the constraint
\begin{eqnarray}
\frac{a}{M_{D_0/D^*/D_1}^4}+\frac{b}{M_{D_0/D^*/D_1}^2}+\frac{c}{s_0}&=&0 \, .
\label{constraint}
\label{aconst}
\end{eqnarray}
Note that in Eq. (\ref{bornconst}) the phenomenological side of 
scattering amplitude for $q_{\mu}\neq 0$ is
not exactly equal to Born term but there are contributions from
other terms. However, for $\omega\rightarrow 0$, $T_N$ on left should
be equal to $T^{Born}$ on right side of Eq. (\ref{bornconst})
and this requirement results in constraint given in Eq. (\ref{constraint}).
As we shall discuss below the constraint
 (\ref{constraint}) help in eliminating the parameter $c$
 and scattering amplitude will be function of parameters $a$ and $b$ only.
The Born terms to be used in equation (\ref{bornconst}) for
 scalar, vector and axial-vector mesons are given by
following relations \cite{wang1,wang2}
 \begin{eqnarray}
 T_{D_0N}^{\rm
Born}(\omega,{\bf0})&=&\frac{2f_{D_0}^2M_{D_0}^2M_N(M_H-M_N)g_{D_0NH}^2}
{\left[\omega^2-(M_H-M_N)^2\right]\left[\omega^2-M_{D_0}^2\right]^2}\,,\nonumber \\
T_{D^*N}^{\rm Born}(\omega,{\bf0})&=&\frac{2f_{D^*}^2M_{D^*}^2M_N(M_H+M_N)g_{D^*NH}^2}
{\left[\omega^2-(M_H+M_N)^2\right]\left[\omega^2-M_{D^*}^2\right]^2}\,, \nonumber \\
T_{D_1N}^{\rm Born}(\omega,{\bf0})&=&\frac{2f_{D_1}^2M_{D_1}^2M_N(M_H-M_N)g_{D_1NH}^2}
{\left[\omega^2-(M_H-M_N)^2\right]\left[\omega^2-M_{D_1}^2\right]^2}\,.
\end{eqnarray}
In the above equations  $g_{D_0NH}$, $g_{D^{\star}NH}$ and $g_{D_1NH}$ are the coupling constants.
 $M_H$ is the the mass of the hadron e.g. corresponding to charm mesons
 we have $\Lambda_c$ and $\Sigma_c$ whereas corresponding to bottom mesons
 we have the hadrons $\Lambda_b$ and $\Sigma_b$.
Corresponding to charm mesons we take the average value of the masses of $M_{\Lambda_c}$ and $M_{\Sigma_c}$ and 
 is equal to $2.4$ GeV. For the case of mesons having bottom quark, $b$, we
 consider the average value of masses of $\Lambda_b$ and $\Sigma_b$ and it is equal to 5.7 GeV.

Now we write the equation for the Borel transformation of the scattering matrix on the
phenomenological side and equate that to the Borel transformation of the scattering matrix for the 
operator expansion side. For the scalar meson,  $D_{0}$,  the Borel transformation equation is written as \cite{wang1},
\begin{eqnarray}
&& a \left\{\frac{1}{M^2}e^{-\frac{M_{D_0}^2}{M^2}}-\frac{s_0}{M_{D_0}^4}e^{-\frac{s_0}{M^2}}\right\}
+b \left\{e^{-\frac{M_{D_0}^2}{M^2}}-\frac{s_0}{M_{D_0}^2}e^{-\frac{s_0}{M^2}}\right\}\nonumber\\
&+& A \left[\frac{1}{(M_H-M_N)^2-M_{D_0}^2}-\frac{1}{M^2}\right]
e^{-\frac{M_{D_0}^2}{M^2}} 
- \frac{A}{(M_H-M_N)^2-M_{D_0}^2}e^{-\frac{(M_H-M_N)^2}{M^2}}\nonumber\\
&=&\left\{\frac{m_c\langle\bar{q}q\rangle_N}{2}\right.
\left.-\langle q^\dag i D_0q\rangle_N+\frac{m_c^2\langle q^\dag i D_0q\rangle_N}{M^2}\right\}e^{-\frac{m_c^2}{M^2}}\nonumber\\
&&-\left\{\frac{2m_c\langle \bar{q} i D_0 i D_0q\rangle_N}{M^2}-\frac{m_c^3\langle \bar{q} i D_0 i D_0q\rangle_N}{M^4}\right\}e^{-\frac{m_c^2}{M^2}}\nonumber\\
&&+\frac{1}{16}\langle\frac{\alpha_sGG}{\pi}\rangle_N\int_0^1 dx \left(1+\frac{\widetilde{m}_c^2}{M^2}\right)e^{-\frac{\widetilde{m}_c^2}{M^2}}
-\frac{1}{48M^4}\langle\frac{\alpha_sGG}{\pi}\rangle_N\int_0^1 dx \frac{1-x}{x}\widetilde{m}_c^4e^{-\frac{\widetilde{m}_c^2}{M^2}}\,,
\label{qcdsumd0}
\end{eqnarray}
where, $A = \frac{2f_{D_0}^2M_{D_0}^2M_N(M_H-M_N)g_{D_0NH}^2}{(M_H-M_N)^2-M_{D_0}^2}$.

Note that in equation (\ref{qcdsumd0}) we have two unknown parameters $a$ and $b$.
We differentiate  equation (\ref{qcdsumd0})  w.r.t.
$\frac{1}{M^{2}}$  so that we could have two equations and two unknowns. By solving those two coupled equations we will be able to get the
values of parameters $a$ and $b$. Same procedure will be applied to obtain the values of parameters 
$a$ and $b$ corresponding to vector and axial-vector mesons.
For vector meson, $D^{*}$, the Borel transformation equation is given by \cite{wang2},
\begin{eqnarray}
&& a \left\{\frac{1}{M^2}e^{-\frac{M_{D^*}^2}{M^2}}-\frac{s_0}{M_{D^*}^4}e^{-\frac{s_0}{M^2}}\right\} 
+b \left\{e^{-\frac{M_{D^*}^2}{M^2}}-\frac{s_0}{M_{D^*}^2}e^{-\frac{s_0}{M^2}}\right\}\nonumber\\
&+& B  \left[\frac{1}{(M_H+M_N)^2-M_{D^*}^2}-\frac{1}{M^2}\right]
e^{-\frac{M_{D^*}^2}{M^2}}
-\frac{B}{(M_H+M_N)^2-M_{D^*}^2}e^{-\frac{(M_H+M_N)^2}{M^2}}\nonumber\\
&=&\left\{-\frac{m_c\langle\bar{q}q\rangle_N}{2}\right.
\left.-\frac{2\langle q^\dag i D_0q\rangle_N}{3}+\frac{m_c^2\langle q^\dag i D_0q\rangle_N}{M^2}
\right\}e^{-\frac{m_c^2}{M^2}}
+\frac{m_c\langle\bar{q}g_s\sigma Gq\rangle_N}{3M^2}e^{-\frac{m_c^2}{M^2}}\nonumber\\
&+&\left\{\frac{8m_c\langle \bar{q} i D_0 i D_0q\rangle_N}{3M^2}-\frac{m_c^3\langle \bar{q} i D_0 i D_0q\rangle_N}{M^4}\right\}e^{-\frac{m_c^2}{M^2}}\nonumber\\
&&-\frac{1}{24}\langle\frac{\alpha_sGG}{\pi}\rangle_N\int_0^1 dx \left(1+\frac{\widetilde{m}_c^2}{2M^2}\right)e^{-\frac{\widetilde{m}_c^2}{M^2}}\nonumber\\
&+&\frac{1}{48M^2}\langle\frac{\alpha_sGG}{\pi}\rangle_N\int_0^1 dx\frac{1-x}{x}\left(\widetilde{m}_c^2-\frac{\widetilde{m}_c^4}{M^2}\right)e^{-\frac{\widetilde{m}_c^2}{M^2}}\,.
\label{qcdsumdst}
\end{eqnarray}
where, $B = \frac{2f_{D^*}^2M_{D^*}^2M_N(M_H+M_N)g_{D^*NH}^2}{(M_H+M_N)^2-M_{D^*}^2}$.
For the axial-vector mesons, $D_{1}$, the Borel transformation equation is given by \cite{wang2},
\begin{eqnarray}
&& a \left\{\frac{1}{M^2}e^{-\frac{M_{D_1}^2}{M^2}}-\frac{s_0}{M_{D_1}^4}e^{-\frac{s_0}{M^2}}\right\}
+ b \left\{e^{-\frac{M_{D_1}^2}{M^2}}-\frac{s_0}{M_{D_1}^2}e^{-\frac{s_0}{M^2}}\right\}\nonumber\\
&+& C \left[\frac{1}{(M_H-M_N)^2-M_{D_1}^2}-\frac{1}{M^2}\right]
e^{-\frac{M_{D_1}^2}{M^2}}
-\frac{C}{(M_H-M_N)^2-M_{D_1}^2}e^{-\frac{(M_H-M_N)^2}{M^2}}\nonumber\\
&=&\left\{\frac{m_c\langle\bar{q}q\rangle_N}{2}\right.
\left.-\frac{2\langle q^\dag i D_0q\rangle_N}{3}+\frac{m_c^2\langle q^\dag i D_0q\rangle_N}{M^2}\right\}e^{-\frac{m_c^2}{M^2}}
-\frac{m_c\langle\bar{q}g_s\sigma Gq\rangle_N}{3M^2}e^{-\frac{m_c^2}{M^2}}\nonumber\\
&-&\left\{\frac{8m_c\langle \bar{q} i D_0 i D_0q\rangle_N}{3M^2}-\frac{m_c^3\langle \bar{q} i D_0 i D_0q\rangle_N}{M^4}\right\}e^{-\frac{m_c^2}{M^2}}\nonumber\\
&&-\frac{1}{24}\langle\frac{\alpha_sGG}{\pi}\rangle_N\int_0^1 dx \left(1+\frac{\widetilde{m}_c^2}{2M^2}\right)e^{-\frac{\widetilde{m}_c^2}{M^2}}\nonumber\\
&+&\frac{1}{48M^2}\langle\frac{\alpha_sGG}{\pi}\rangle_N\int_0^1 dx\frac{1-x}{x}\left(\widetilde{m}_c^2-\frac{\widetilde{m}_c^4}{M^2}\right)e^{-\frac{\widetilde{m}_c^2}{M^2}}\,,
\label{qcdsumd1}
\end{eqnarray}
where, $C = \frac{2f_{D_1}^2M_{D_1}^2M_N(M_H-M_N)g_{D_1NH}^2}{(M_H-M_N)^2-M_{D_1}^2}$.
In the above equations  $\widetilde{m}_c^2=\frac{m_c^2}{x}$.

As discussed earlier, in determining the properties of hadrons from QCD sum rules, we shall use the 
values of quark and gluon condensates as calculated using chiral SU(3) model.
Any operator $\mathcal{O}$ on OPE side can be written as \cite{koike1,zscho1,kwon1},
\begin{eqnarray}
\mathcal{O}_{\rho_{B}} &=&\mathcal{O}_{vacuum} +
4\int\frac{d^{3}p}{(2\pi)^{3} 2 E_{p}}n_{F}\left\langle N(p)\vert \mathcal{O}\vert N(p) \right\rangle+
3\int\frac{d^{3}k}{(2\pi)^{3} 2 E_{k}}n_{B}\left\langle \pi(k)\vert \mathcal{O}\vert\pi(k) \right\rangle \nonumber\\
& =& \mathcal{O}_{vacuum} + \frac{\rho_B}{2M_N}\mathcal{O}_{N} + 
\mathcal{O}_{P.B.}.
\label{operator1}
\end{eqnarray}
In above equation, $\mathcal{O}_{\rho_{B}}$ , gives
us the expectation value of the operator at finite baryonic density. 
The term $\mathcal{O}_{vacuum}$ stands for the vacuum expectation value of the
operator, $\mathcal{O}_{N}$ gives us the nucleon expectation value
of the operator and $\mathcal{O}_{P.B.}$ denotes
the contribution from the pion bath at finite temperature.
$n_{B} = \left[ e^{E_{k}/T} - 1 \right]^{-1}$ and 
$n_{F} =  \left[ e^{\left( E_{p}-\mu_{N}\right) /T} - 1 \right]^{-1}$ 
are the thermal Boson and Fermion distribution functions.
Within the chiral SU(3) model the quark and gluon condensates
can be expressed in terms of scalar fields $\sigma$, $\zeta$ and 
$\chi$.
As discussed earlier, the finite temperature effects in the present investigation will be evaluated through the scalar fields and therefore
contribution of third term will not be considered. However,
for completeness we shall compare the temperature dependence of
scalar quark and scalar gluon condensates at zero baryon density as evaluated in the
present work with the situation when the temperature
dependence is evaluated using only pion bath contribution \cite{kwon1}.
Thus within chiral SU(3) model, we can find the values of
 $\mathcal{O}_{\rho_{B}}$ at finite density of the nuclear medium and hence can find
 $\mathcal{O}_{N}$ using
\begin{equation}
\mathcal{O}_{N} = \left[ \mathcal{O}_{\rho_{B}}  - \mathcal{O}_{vacuum}\right] \frac{2M_N}{\rho_B}.
\label{condexp}
\end{equation}
The quark condensate, $\bar{q}q$, 
can be extracted from the explicit symmetry breaking term 
of the Lagrangian density and is given by,
 \begin{eqnarray}
&&\sum_i m_i \bar {q_i} q_i = - {\cal L} _{SB} 
 = \left( \frac {\chi}{\chi_{0}}\right)^{2} 
\left[ m_{\pi}^{2} 
f_{\pi} \sigma
+ \big( \sqrt {2} m_{k}^{2}f_{k} - \frac {1}{\sqrt {2}} 
m_{\pi}^{2} f_{\pi} \big) \zeta \right]. 
\end{eqnarray}
 In our present investigation of hadron properties,
 we are interested in light quark condensates, $\bar{u}u$ and $\bar{d}d$, which are proportional
 to the non-strange scalar field $\sigma$ within chiral $SU(3)$ model.
 Considering equal mass of light quarks, $u$ and $d$ i.e. 
 $m_{u} = m_{d} = m_{q} = 0.006$ GeV, we can write,
 \begin{equation}
\left\langle \bar{q}q\right\rangle _{\rho_{B}} 
= \frac{1}{2m_{q}}\left( \frac {\chi}{\chi_{0}}\right)^{2} 
\left[ m_{\pi}^{2} 
f_{\pi} \sigma \right] .
\label{quarkcondmodel}
\end{equation}
The condensate $\langle\bar{q}g_s\sigma Gq\rangle_{\rho_B}$ is given by the equation \cite{qcdThomas},
\begin{equation}
\langle\bar{q}g_s\sigma Gq\rangle_{\rho_B} = \lambda^{2}\left\langle \bar{q}q \right\rangle_{\rho_{B}} + 3.0 GeV^{2}\rho_{B}.
\label{quarkcond2}
\end{equation}
Also we write \cite{qcdThomas},
\begin{equation}
\langle \bar{q} i D_0 i D_0q\rangle_{\rho_B} + \frac{1}{8}\langle\bar{q}g_s\sigma Gq\rangle_{\rho_B} =  0.3 GeV^{2}\rho_{B}.
\label{quarkcond3}
\end{equation}
As discussed above the quark condensate, $ \left\langle \bar{q}q\right\rangle _{\rho_{B}} $,
can be calculated within the chiral SU(3) model.
 This value of $ \left\langle \bar{q}q\right\rangle _{\rho_{B}} $
can be used through equations (\ref{quarkcond2}) and (\ref{quarkcond3}) to
calculate the value of condensates $\langle\bar{q}g_s\sigma Gq\rangle_{\rho_B}$ and
 $\langle \bar{q} i D_0 i D_0q\rangle_{\rho_B}$ within chiral SU(3) model.
 The value of condensate $\langle q^\dag i D_0q\rangle$ is equal to  $0.18 GeV^2 \rho_{B}$ \cite{qcdThomas}.
 
At finite temperature and zero baryon density we can write the expectation values of  quark condensates and scalar gluon condensates as \cite{kwon1},
\begin{equation}
 \left\langle \bar{q}q\right\rangle _{T } = \left\langle \bar{q}q\right\rangle _{vacuum}
 \left[ 1- \frac{T^{2}}{8f_{\pi}^{2}}B_{1}\left(\frac{m_{\pi}}{T} \right)\right],
 \label{quark_pion} 
\end{equation}
and
\begin{equation}
\left\langle  \frac{\alpha_{s}}{\pi} {G^a}_{\mu\nu} {G^a}^{\mu\nu} 
\right\rangle_{T} = \left\langle  \frac{\alpha_{s}}{\pi} {G^a}_{\mu\nu} {G^a}^{\mu\nu} 
\right\rangle_{vacuum} - \frac{1}{9}m_{\pi}^{2}T^{2}B_{1}\left(\frac{m_{\pi}}{T} \right).
 \label{gluon_pion} 
\end{equation}
respectively.
 In equations (\ref{quark_pion}) and (\ref{gluon_pion}) $B_{1}(x) = \frac{6}{\pi^{2}}\int_{x}^{\infty}dy \frac{\sqrt{y^{2}-x^{2}}}{e^{y} -1}$.
From equations (\ref{quark_pion}) and (\ref{gluon_pion}) 
we observe that the contribution from pion bath to the expectation values
of operator arises only at finite temperature.
Note that in equations (\ref{qcdsumd0}), (\ref{qcdsumdst}) and
 (\ref{qcdsumd1}) we need the nucleon expectation values of
 various condensates which can be evaluated
  in general using equation(\ref{condexp}). In equations (\ref{chiglu}), (\ref{quarkcondmodel}),
  (\ref{quarkcond2}) and (\ref{quarkcond3}) the values of condensates
  are given at finite value of baryonic density.
  To find the corresponding nucleon expectation values of various condensates
  we use the values of condensates at finite baryonic density
      from equations (\ref{chiglu}), (\ref{quarkcondmodel}),
  (\ref{quarkcond2}) and (\ref{quarkcond3}) 
  in equation(\ref{condexp}). The equation (\ref{condexp}) is then
  further used in equations (\ref{qcdsumd0}), (\ref{qcdsumdst}) and
 (\ref{qcdsumd1}) for calculation of medium modifications of $D$ mesons.
 
 It may be noted that the QCD sum rules for the evaluation of in-medium properties of scalar mesons, $B_0$,
  vector mesons, $B^*$ and axial vector mesons $B_1$, can be written by replacing 
  masses of charmed mesons $D_0$, $D^*$ and $D_1$, by  
  corresponding masses of  bottom mesons $B_0$, $B^*$ and $B_1$ 
  in equations (\ref{qcdsumd0}), (\ref{qcdsumdst}) and (\ref{qcdsumd1}) respectively. 
   Also the bare charm quark mass, $m_c$ will be replaced by the mass of bottom quark, $m_b$.
   
 \section{Results And Discussions}
 
 In this section we shall present the results of our investigation 
 of in-medium properties of scalar $\left(D_0, B_0 \right)$,  
 vector $\left(D^*, B^* \right)$ and axial vector mesons $\left(D_1, B_1 \right)$.
 The nuclear matter saturation density 
used in the present investigation is $0.15$ fm$^{-3}$.
The values of various coupling constants $g_{ D_0 N \Lambda_c}\approx g_{ D_0 N \Sigma_c}
\approx g_{ B_0 N \Lambda_b}\approx g_{B_0 N \Sigma_b}$ are approximated to 6.74 \cite{wang1}.
The coupling constants $g_{D^*N \Lambda_c}\approx g_{D^*N \Sigma_c}
\approx g_{D_1N\Lambda_c}\approx g_{D_1N\Sigma_c}\approx g_{B^*N\Lambda_b}\approx g_{B^*N\Sigma_b}
\approx g_{B_1N\Lambda_b}\approx g_{B_1N\Sigma_b}$ are approximated to 3.86 \cite{wang2}.
The masses of mesons $M_{D_0}$, $M_{B_0}$, $M_{D^*}$, $M_{B^*}$,
$M_{D_1}$ and  $M_{B_1}$ to be used in present investigation are 
2.355, 5.74, 2.01, 5.325, 2.42 and  5.75
GeV respectively. 
The values of decay constants $f_{D_0}$, $f_{B_0}$, $f_{D^*}$, $f_{B^*}$,
$f_{D_1}$ are $f_{B_1}$ are 0.334, 0.28, 0.270, 0.195, 0.305 
and 0.255 GeV respectively.
The values of threshold parameters, $s^0$, corresponding to 
$D_0$, $B_0$, $D^*$, $B^*$,
$D_1$ and  $B_1$ mesons are
8, 39, 6.5, 35, 8.5 and 39 GeV$^2$ respectively \cite{wang1,wang2}.
As discussed earlier the mass-shift of
scalar, vector and axial-vector mesons
is calculated through the parameter $a$ which is related to scattering length
through equations (\ref{a2}) to (\ref{massshift}).
 This parameter $a$, for example, for $D^{0}$ is 
calculated by solving the coupled equations as discussed after Eq. (\ref{qcdsumd0})
and is subjected to the medium modifications
through the medium dependence of condensates.
The medium dependence of condensates is further evaluated through the
scalar fields $\sigma$, $\zeta$ and $\chi$.
The various coupling constants listed above, the decay constants of $D$ and $B$ mesons,
the threshold parameter $s_0$ are not subjected to the
medium modifications. In the present work we shall show the variation of
mass shift as a function of squared Borel mass parameter, M.
The Borel window is chosen such that there is almost no change in the
mass of $D$ and $B$ mesons  w.r.t variation in Borel mass parameter.
For the charmed scalar, $D_0$, vector, $D^{\star}$,
and axial-vector, $D_1$  mesons  Borel windows are 
found to be $(6.1-7.4)$, $(4.5-5.4)$ and $(6.5-7.6)$ GeV$^{2}$
respectively. The Borel windows for bottom scalar, $B_0$, vector, $B^{\star}$,
and axial-vector, $B_1$  mesons are 
 $(33-39)$, $(22-24)$ and $(34-37)$ GeV$^{2}$
respectively.

In the present work we are studying the 
in-medium masses of scalar, vector and axial vector mesons
using QCD sum rules and chiral SU(3) model. Since we are evaluating the quark and gluon condensates 
within the chiral SU(3) model through the modification of scalar fields $\sigma$ and $\zeta$ and the scalar
dilaton field $\chi$, so first we shall discuss in short the effect of temperature and density of the medium 
on the values of scalar fields $\sigma$ and $\zeta$ and the scalar
dilaton field $\chi$. 
We observe that as a function of density of the nuclear medium the magnitude of 
scalar fields decreases. For example at nuclear saturation density, $\rho_{B} = \rho_{0}$,
the drop in magnitude of the scalar fields $\sigma$ and $\zeta$ and the dilaton field $\chi$
is observed to be $33.49, 9.54$ and $3.39$ MeV respectively. At baryon densities, $4\rho_{0}$, these
values changes to $63.41, 14.51$ and $13.06$ MeV respectively.

At zero baryon density, we observe 
that the magnitude of the scalar fields $\sigma$ and $\zeta$
and the dilaton field $\chi$ 
decreases with increase in the temperature. However, the change
in the values of scalar fields with temperature
of the medium is observed to be very small. The reason for the
non zero values of scalar fields at finite temperature and zero baryon density of the 
medium is the formation of baryon-antibaryon pairs \cite{kristof1,frunstl1}.
At finite baryon densities, the magnitude of the scalar fields 
increases with increase in the temperature of symmetric nuclear medium.
This also leads to increase in the masses of the nucleons with the 
temperature of the nuclear medium for finite baryon densities \cite{amarvind}.
At nuclear saturation density, $\rho_{0}$, the magnitude of the
scalar fields, $\sigma$ and $\zeta$
and the dilaton field $\chi$ increases by $5.47, 1.28$ and
 $0.96$ MeV respectively as we move from T = 0 to T = 150
MeV respectively.

\begin{figure}
 \resizebox{1.0 \textwidth}{!}{%
 \includegraphics{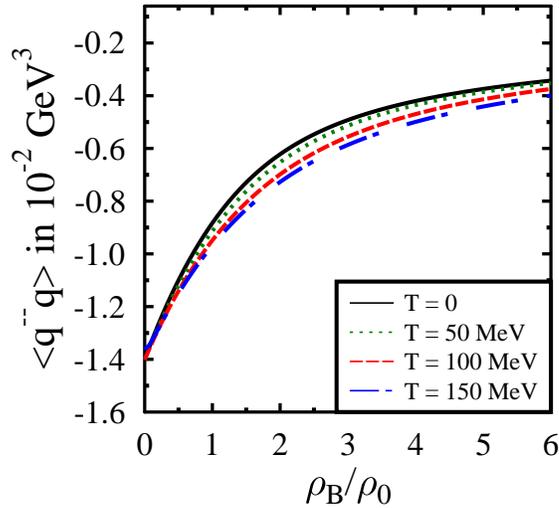}
}
\caption{(Color online)
The light quark condensate $\bar{q}q$ plotted as a function of
density of the nuclear medium, in units of nuclear saturation density,
for different values of temperatures ($T = 0, 50, 100$ and $150$ MeV).}
\label{ltqcondensates}
\end{figure}

\begin{figure}
 \resizebox{1.0\textwidth}{!}{%
 \includegraphics{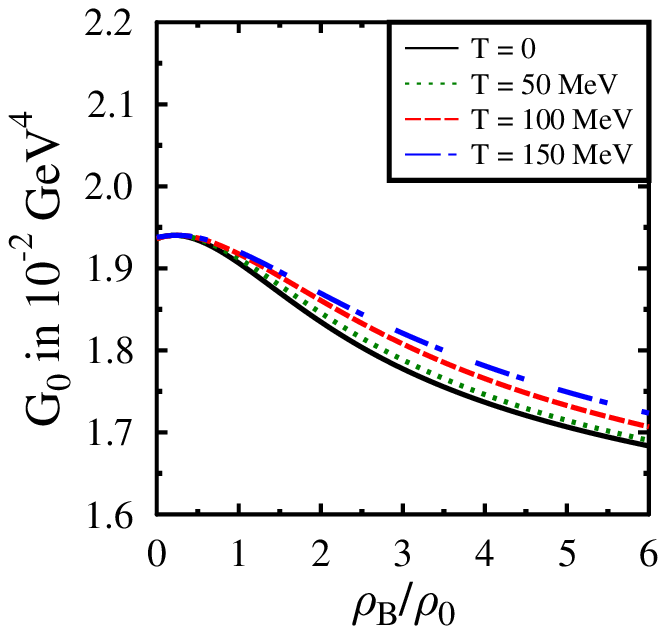}
}  
\caption{(Color online)
The scalar gluon condensate $G_{0}$  plotted as a function of
density of the nuclear medium, in units of nuclear saturation density,
for different values of temperatures ($T = 0, 50, 100$ and $150$ MeV).}
\label{sgcondensates}
\end{figure}

In figures (\ref{ltqcondensates}) and (\ref{sgcondensates}) we show the
variation of the light scalar quark condensate 
$\bar{q}{q}$, given by equation (\ref{quarkcondmodel}) and
the scalar gluon condensate, 
$G_{0} = \left\langle  \frac{\alpha_{s}}{\pi} {G^a}_{\mu\nu} {G^a}^{\mu\nu} 
\right\rangle$, given by equation
(\ref{chiglu}), respectively as a
function of density of the symmetric nuclear medium. We show the results for temperatures,
$T = 0, 50, 100$ and 150 MeV respectively.
 From equation 
(\ref{quarkcondmodel}) we observe that the
value of the scalar quark condensate $\bar{q}q$ is
directly proportional to the scalar-isoscalar field, $\sigma$.
Therefore, the behavior of the $\bar{q}q$ as a 
function of temperature and density of the nuclear
 medium will be same as that of $\sigma$ field.
 For given value of temperature of the nuclear medium the magnitude of the
 light quark condensate decreases with increase in the density.
 For example, at T = 0, the values of light quark condensate are observed to be
 $-0.8829 \times 10^{-2}$ and $-0.4200 \times 10^{-2}$ GeV$^{3}$
  at baryon densities, $\rho_{B} = \rho_{0}$ and
 $4\rho_{0}$ respectively. At temperature, T = 150 MeV, 
 these values of quark condensates changes 
 to $-0.9681 \times 10^{-2}$ and $-0.4993 \times 10^{-2}$ GeV$^{3}$
 respectively.
At zero baryon density the magnitude of the $\bar{q}q$ 
decreases with increase in the temperature of the nuclear medium.
At $\rho_{B}=0$, the values of $\bar{q}q$ are observed to be 
$-1.4014 \times 10^{-2}$, $-1.4014 \times 10^{-2}$, $-1.4006 \times 10^{-2}$
and $-1.3628 \times 10^{-2}$ GeV$^{3}$ at 
 temperatures T = 0, 50, 100 and 150 MeV respectively.
 
 From figure (\ref{sgcondensates}), we observe that
the values of the scalar gluon condensates decreases with increase in the density of the nuclear 
medium.
At baryon density,
 $\rho_B=\rho_0$, the values of $G_0$ are observed to be 
1.90646 $\times 10^{-2}$ GeV$^4$, 1.9119 $\times 10^{-2}$ GeV$^4$, 1.91755 $\times 10^{-2}$ GeV$^4$, 
 and 1.92 $\times 10^{-2}$ GeV$^4$  
for temperatures, T = 0, 50, 100 and 150 MeV respectively. For the same
values of the temperature, in the absence of finite quark masses,
the values of $G_0$ are observed to be 2.269 $\times 10^{-2}$ GeV$^4$, 
2.2771$\times 10^{-2}$ GeV$^4$, 2.2857 $\times 10^{-2}$ GeV$^4$ and 2.29 $\times 10^{-2}$ GeV$^4$
 for $\rho_B=\rho_0$.
For baryon density $\rho_B=4\rho_0$,
the values of $G_0$ are given as 1.7367 $\times 10^{-2}$ GeV$^4$ 
(2.06 $\times 10^{-2}$ GeV$^4$), 1.74612 $\times 10^{-2}$ GeV$^4$ 
(2.0713 $\times 10^{-2}$ GeV$^4$), 1.7656 $\times 10^{-2}$ GeV$^4$ 
(2.094 $\times 10^{-2}$ GeV$^4$) and  1.78 $\times 10^{-2}$ GeV$^4$ 
(2.112 $\times 10^{-2}$ GeV$^4$)  for values of temperature, T = 0, 50, 100 and 
150  MeV respectively, for the cases of the finite (zero) quark 
masses in the trace anomaly.

 It may be noted that in the above discussion the finite temperature
 effects of the nuclear medium on the values of quark and
 gluon condensates are evaluated through the
 temperature dependence of scalar fields $\sigma$, $\zeta$
 and the dilaton field $\chi$. In literature the scalar
 quark and gluon condensates at finite temperature
 are evaluated due to contribution from pion bath using equations
 (\ref{quark_pion}) and (\ref{gluon_pion}) respectively \cite{kwon1}.
 Using $m_{q}\left\langle  \bar{q}q\right\rangle   _{0}$ = - (0.11 GeV) $^{4}$
and $\left\langle \frac{\alpha_{s}}{\pi}G^{2}\right\rangle $ = 0.005 GeV$^{4}$ \cite{kwon1}
 we calculate the quark and gluon condensates at finite
 temperature and zero baryon density using equations
 (\ref{quark_pion}) and (\ref{gluon_pion}) respectively.
 The values of scalar quark condensates are observed to be
 $-1.824\times 10^{-2}$, $-1.751\times 10^{-2}$ and $-1.566 \times 10^{-2}$ GeV$^{3}$ at temperatures T = 50, 100 and 150 MeV respectively.
 These values of scalar quark condensates  can be compared
 to the values $-1.4014\times 10^{-2}$, $-1.4006 \times 10^{-2}$
and $-1.3628 \times 10^{-2}$ GeV$^{3}$ at T = 50, 100 and 150 MeV
 calculated in our present approach where finite
  temperature effects were considered through 
  scalar fields $\sigma$, $\zeta$ and $\chi$.
 Similarly, the values of scalar gluon condensates
 calculated using equation (\ref{gluon_pion}) are  found to
 be $0.5\times 10^{-2}$, $0.4994 \times 10^{-2}$ and $0.497 \times 10^{-2}$
  GeV$^{4}$ at T = 50, 100 and 150 MeV respectively.
  However, for zero baryon density, using the chiral SU(3) model 
  the values of scalar gluon condensates
   for finite (zero) quark mass term,
  are observed to be
    $1.94264\times 10^{-2}$ ($2.34550 \times 10^{-2}$), $1.94269\times 10^{-2}$ 
    ($2.3454 \times 10^{-2}$)
     and $1.94456\times 10^{-2}$ ($2.3437 \times 10^{-2}$) GeV$^{4}$
     at temperatures T = 50, 100 and 150 MeV
  respectively.
  From above discussion we observe that as a function of temperature there
 are very small variations in the values of scalar quark and 
 gluon condensates in the nuclear medium. This observation agree
 e.g. with ref. \cite{chiral1}
 where gluon and chiral condensates were studied at finite temperature with an effective
Lagrangian of pseudoscalar mesons coupled to a scalar glueball.
The gluon condensates were found to be very stable up to temperatures of 200 MeV, where
the chiral sector of the theory reaches its limit of validity \cite{chiral1}.
Actually scalar quark and gluon condensates are found to vary effectively with temperature
 above critical temperature but in hadronic medium, for zero baryon density,
  these are not much sensitive to temperature
 effects \cite{glu22,glu23}.

Note that in the above discussion of scalar gluon and quark
condensates, calculated using equations (\ref{chiglu}) and
(\ref{quarkcondmodel}) respectively,
 we considered the vacuum values of decay constants
$f_{\pi}$ and $f_{K}$ as well as masses $m_{\pi}$ and $m_K$ of 
pions and kaons respectively. Now we shall discuss the effect of medium
modified values of $f_{\pi}$, $f_{K}$, $m_{\pi}$ and $m_K$
on the scalar condensates. In the chiral effective model the
pion and kaon decay constants are related to the
scalar fields $\sigma$ and $\zeta$ through relations
\cite{paper3,amarvind}
\begin{equation}
f_{\pi} = -\sigma
\label{pdecay}
\end{equation}
and 
\begin{equation}
f_{K} = -\frac{1}{2}\left( \sigma +\sqrt{2}\zeta \right) 
\label{kdecay}
\end{equation}
respectively.
From the above relations it is clear that the medium dependence
 of scalar fields $\sigma$ and $\zeta$ can be used to study
  the density and temperature 
  dependence of decay constants of pions and kaons. 
  For example, using equation (\ref{pdecay}), at zero temperature
  the values of pion decay constant are
  observed to be $93.3$, $59.75$ and $42.99$ MeV at
  baryonic densities $\rho_B$ = 0, $\rho_0$
  and 2$\rho_0$ respectively.
  The values of kaons decay constants,  using equation (\ref{kdecay}), are observed to be
  $122$, $98.6$ and $87.93$ MeV at $\rho_B$ = 0, $\rho_0$
  and 2$\rho_0$ respectively. 
  In reference \citep{pionmass1}  
  the masses of pions were calculated in linear density 
  approximation using the chiral perturbation theory.
  At baryon density $\rho_B$ = $\rho_0$ and zero temperature
  the mass of pion changes from the vacuum value $139$ MeV to
  $145.8$ MeV. 
  The properties of kaons and antikaons had been
investigated in the literature using the chiral model \cite{isoamss2} 
and coupled channel approach 
 \cite{Lutz98,Oset00,Lutz02,lauran}.
In the present work, to study the effect of medium modified 
masses of kaons on the values of 
quark and gluon condensates we shall use the chiral SU(3) model \cite{isoamss2}.
In the nuclear medium the kaons feel repulsive interactions
and their in-medium mass increases as a function of
density whereas the antikaons feel attractive interactions and
their in-medium masses drop as we move to higher baryon density.
For the present purpose considering the average mass of kaons and antikaons, 
the in-medium mass at temperature $T = 0$ 
is observed to be 494, 488.88 and 466.08 MeV at $\rho_B$ = 0, $\rho_0$ and 
$2\rho_0$ respectively. Taking into account the
above discussed in-medium properties of $f_{\pi}$, $f_K$, $m_{\pi}$
and $m_K$, at
baryonic density $\rho_B$ = $\rho_0$ and zero temperature
 the values of scalar quark and gluon condensates
are observed to be -0.6222 $\times 10^{-2}$ GeV$^3$ 
and 1.9869 $\times 10^{-2}$ GeV$^4$  respectively.
These values of scalar quark and gluon condensates can 
be compared to 
-0.8829 $\times 10^{-2}$ GeV$^3$ 
and 1.9065 $\times 10^{-2}$ GeV$^4$ respectively which were calculated
without the medium modification of 
$f_{\pi}$, $f_K$, $m_{\pi}$
and $m_K$. We conclude that the medium modification of 
$f_{\pi}$, $f_K$, $m_{\pi}$
and $m_K$ causes more decrease in the values of scalar quark and 
gluon condensates at finite baryonic density.

\begin{figure}
 \resizebox{0.8\textwidth}{!}{
 \includegraphics{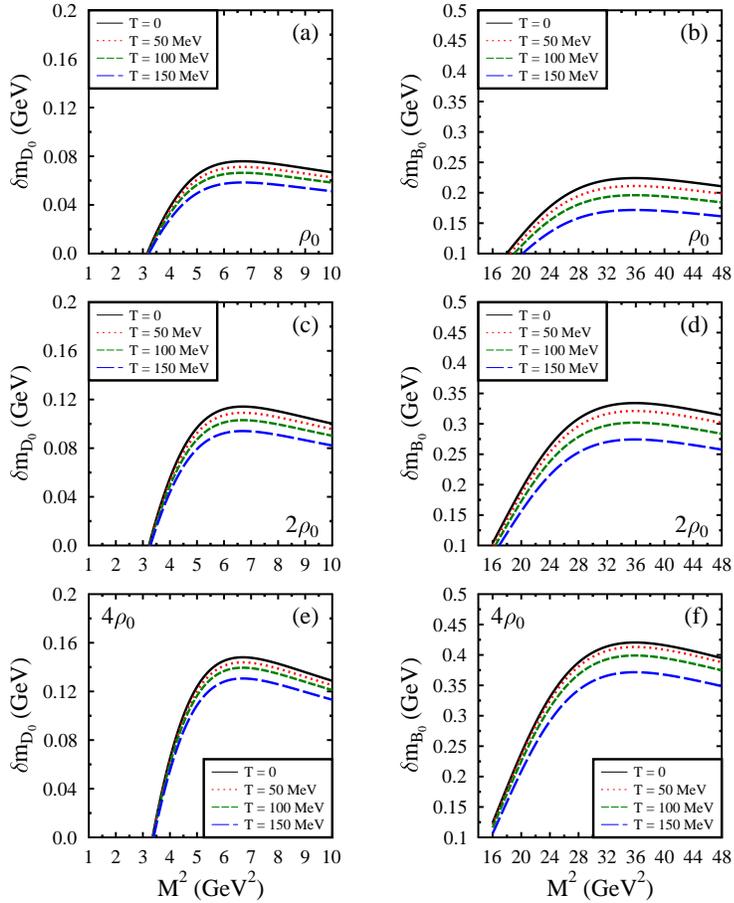}
}  
\caption{(Color online)
Figure shows the variation of the mass shift of Scalar mesons $D_{0}$ (subplots (a), (c), (e)) and $B_{0}$
(subplots (b), (d), (f)) as a function of the squared Borel mass Parameter, $M^2$.
We show the results at baryon densities $\rho_0$, $2\rho_0$ and $4\rho_0$. For each value of 
baryon density the results are shown at temperatures, T = 0, 50, 100 and 150 MeV. Also
the medium modifications of decay constants and masses of pions and kaons are
considered while evaluating the above shown mass-shift of scalar mesons.}
\label{scalarDBfig1}
\end{figure}

Now we shall calculate the in-medium masses of scalar, vector and axial 
vector mesons using the values of condensates from chiral SU(3) model.
In figure (\ref{scalarDBfig1}), the subfigures (a), (c) and (e) show the 
variation of the mass shift of scalar mesons $D_0$ as a function of
square of the Borel mass parameter, $M$,
at nuclear matter densities $\rho_0$,  $2\rho_0$ and $4\rho_0$ respectively.
The subfigures (b), (d) and (f) show the 
variation of the scalar mesons $B_0$ as a function of 
square of the Borel mass parameter, $M$,
at nuclear matter densities $\rho_0$,  $2\rho_0$ and $4\rho_0$ respectively.
In each subplot we have shown the results at 
temperatures T = 0, 50, 100 and 150 MeV.
We observe that for scalar mesons, $D_0$, at temperature, T = 0, the values of mass
shift are found to be 76, 114 and 148 MeV at baryon
densities $\rho_0$,  $2\rho_0$ and $4\rho_0$ respectively.
For temperature T = 50 MeV the values of mass shift are observed to be  71, 109 and 144 MeV at baryon
densities $\rho_0$,  $2\rho_0$ and $4\rho_0$ respectively.
At temperature T = 100 MeV the above values of mass shift are observed to be
66, 103 and 139 MeV, whereas at T = 150 MeV 
 the values of mass shift changes to
58, 94 and 131 MeV at baryon
densities $\rho_0$,  $2\rho_0$ and $4\rho_0$ respectively.
From the above discussion we conclude that for a given value of temperature 
the mass shift of scalar mesons, $D_0$, increases
as a function of density of
the nuclear medium.
On the other hand as a function of temperature of the nuclear medium, for a constant value of density,
the mass shift of scalar mesons, $D_0$, decreases.

As we can see from figure (\ref{scalarDBfig1}) the values of mass shift of scalar $B_0$ mesons
also increases as a function of density of the nuclear medium. At temperature $T = 0$,
the values of mass shift are observed to 224, 334 and 420 MeV at
baryon densities $\rho_0$,  $2\rho_0$ and $4\rho_0$ respectively.
At temperature T = 50 MeV the above values of  mass shift are
found to be 211, 321 and 413 MeV at baryon densities
  $\rho_0$,  $2\rho_0$ and $4\rho_0$ respectively.
 For temperature T = 100 MeV the values of mass shift are found to 
 be 196, 302 and 399 MeV whereas at T = 150 MeV these values of 
 mass shift changes to 171, 274 and 372 MeV at
  $\rho_0$,  $2\rho_0$ and $4\rho_0$ respectively.
  For a given value of nuclear matter density the values of mass shift
   for the $B_0$ mesons are found to decrease
  with increase in the temperature of the medium. 
  
We also calculate the values of scattering lengths of scalar mesons using 
equation (\ref{a2}) for different values of  density and temperature 
of the medium. At temperature T = 0 and baryon densities, $\rho_B$ = $\rho_0$ and $4\rho_0$
 the values of scattering lengths
for scalar mesons $D_0$ ($B_0$) are observed to be 1.42 (5.05) and 0.70 (2.41) fm respectively.
For temperature T = 100 MeV and baryon densities, $\rho_B$ = $\rho_0$ and $4\rho_0$
the values of scattering lengths for  
$D_0$ ($B_0$) are observed to be 1.23 (4.40) and 0.66 (2.28) fm respectively.
We note that the value of scattering length decreases as a
function of density and temperature of the nuclear medium.
As discussed above the scattering lengths are evaluated using equation (\ref{a2}).
In this equation we have the parameter $a$ which is directly proportional to the 
scattering length of mesons. As discussed earlier the value of parameter $a$
is evaluated by solving simultaneously equation (\ref{qcdsumd0})
and the equation obtained by
differentiate  equation (\ref{qcdsumd0})  w.r.t.
$\frac{1}{M^{2}}$. The magnitude of parameter $a$ is found to 
decrease as we move from low  to higher value of baryonic density
or from zero  to finite value of temperature of the medium.
This behavior of parameter $a$ as a function of density and
temperature of the medium results in the similar changes in 
the values of scattering lengths of scalar mesons.

\begin{figure}
 \resizebox{0.8\textwidth}{!}{%
 \includegraphics{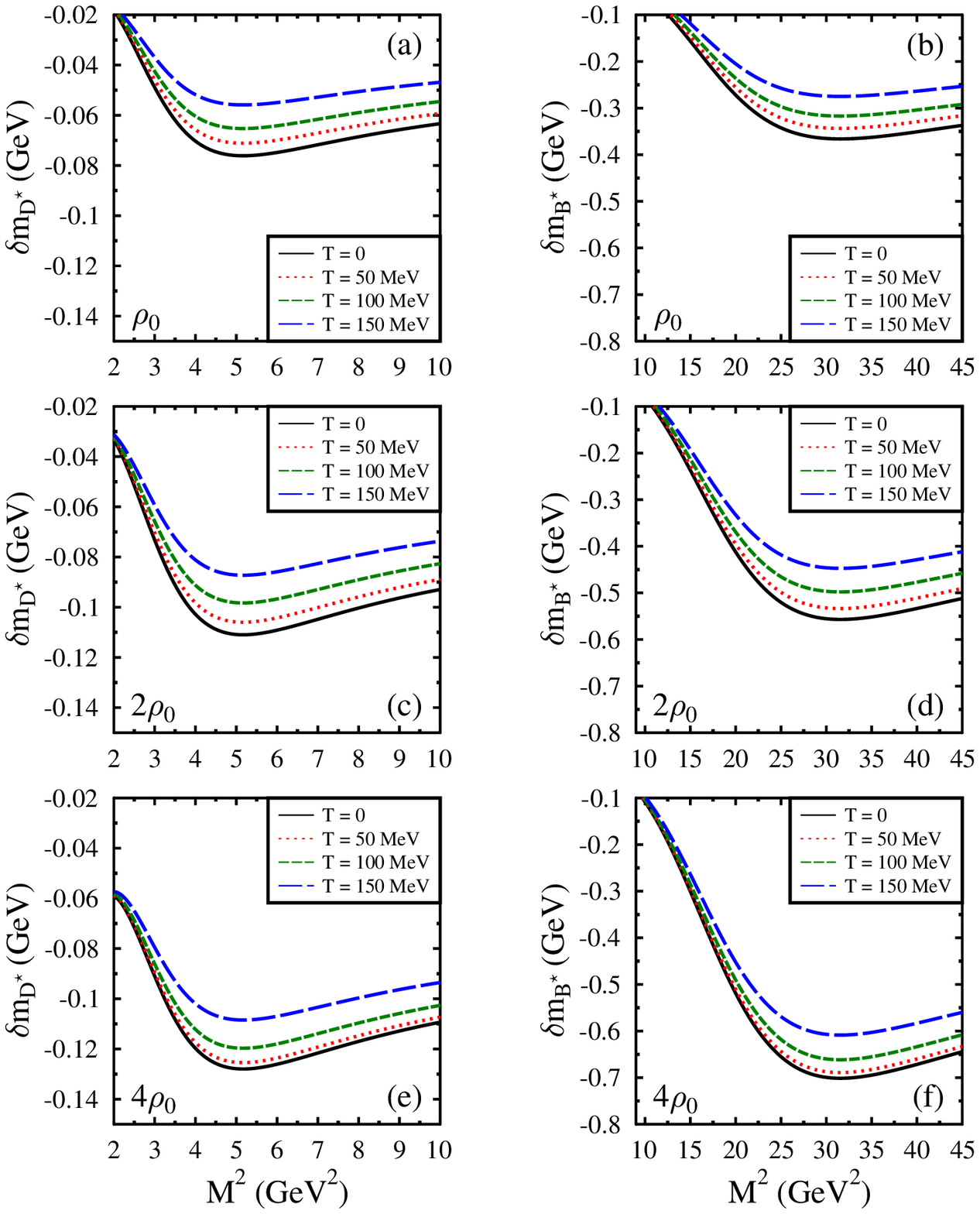}
}
\caption{(Color online)
Figure shows the variation of the mass shift of vector mesons $D^{\star}$ (subplots (a), (c), (e)) and $B^{\star}$
(subplots (b), (d), (f)) as a function of the squared Borel mass Parameter, $M^2$.
We have shown the results at baryon densities $\rho_0$, $2\rho_0$ and $4\rho_0$. For each value of the 
baryon density the results are shown at temperatures, T = 0, 50, 100 and 150 MeV.}
\label{vectorDBfig2}
\end{figure}

Figure (\ref{vectorDBfig2}) shows the variation of the
 mass shift of  vector mesons $D^{\star}$ and $B^{\star}$ as
a function of square of the Borel mass parameter. Here also 
we have shown the results at nuclear matter densities $\rho_0$,  $2\rho_0$ and $4\rho_0$.
We observe that at nuclear matter density $\rho_B = \rho_0$ the values of 
mass shift for vector mesons, $D^{\star}$ are observed to be -76, -71, -65 and $-56$ MeV at
temperatures T = 0, 50, 100 and 150 MeV respectively.
For baryon density $\rho_B = 2\rho_0$ the values of mass shift are found 
to be $-111, -106, -98, $ and $-87$
MeV whereas for  $\rho_B = 4\rho_0$ the values of the mass shift 
changes to $-128, -125, -120$ and $-108$ MeV at temperatures T = 0, 50, 100 and 150 MeV respectively.
Note that for a given value of density the magnitude of the mass shift of vector mesons decreases
as a function of temperature of the nuclear medium. On the other hand as a function of density 
of  the medium the magnitude of the mass shift of the vector mesons, $D^{\star}$ increases.
For nuclear matter saturation density the values of  mass shift for the $B^{\star}$ mesons 
are found to be $-366$, -344, -311 and 
-275 MeV at temperatures T = 0, 50, 100 and 150 MeV respectively.
At baryon density $2\rho_0$ the above values of mass shift changes to 
$-557$, -534, -498 and 
-447 MeV at temperatures T = 0, 50, 100 and 150 MeV respectively
whereas for baryon density $4\rho_0$ the above values of mass shift found to  be
$-701$, -689, -662 and 
-609 MeV at temperatures T = 0, 50, 100 and 150 MeV respectively.
At temperature T = 0 and baryon densities, $\rho_B$ = $\rho_0$ and $4\rho_0$
 the values of scattering lengths
for vector mesons $D^\star$ ($B^\star$) are observed to be -1.31 (-7.72) and -0.54 (-3.58) fm respectively.
For temperature T = 100 MeV abd baryon densities, $\rho_B$ = $\rho_0$ and $4\rho_0$
the values of scattering lengths for  
$D^\star$ ($B^\star$) are observed to be -1.12 (-6.72) and -0.51 (-3.54) fm respectively.
We observe that the magnitude of the scattering lengths of vector mesons decreases
in moving from low to higher value of density or temperature of the
medium.

\begin{figure}
 \resizebox{0.8\textwidth}{!}{%
 \includegraphics{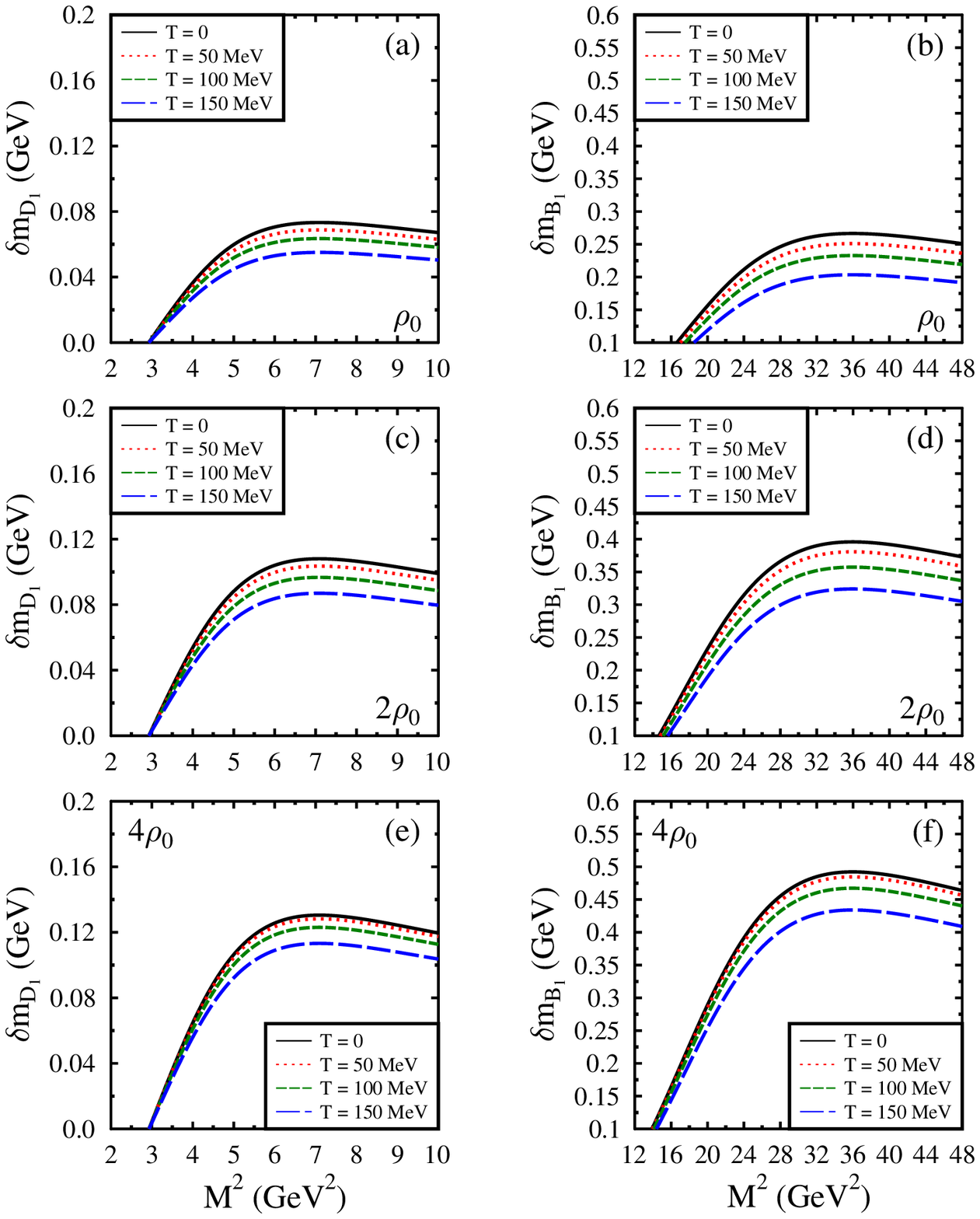}
} 
\caption{(Color online)
Figure shows the variation of the mass shift of axial vector mesons $D_{1}$ (subplots (a), (c), (e)) and $B_{1}$
(subplots (b), (d), (f)) as a function of the squared Borel mass Parameter, $M^2$.
We have shown the results at baryon densities $\rho_0$, $2\rho_0$ and $4\rho_0$. For each value of the 
baryon density the results are shown at temperatures, T = 0, 50, 100 and 150 MeV.}
\label{axialvectorDBfig3}
\end{figure}

In figure (\ref{axialvectorDBfig3}), for  given values of temperatures and densities we have shown the variation of 
 mass shift of  axial-vector mesons $D_{1}$ and $B_{1}$ as
a function of square of the Borel mass parameter.
We observe that at baryon density $\rho_B = \rho_0$,
 the values of mass shift for axial vector meson $D_1$
 are observed to be $73$, 69, 63 and 55 MeV at temperatures, T = 0, 50, 100 and 150 MeV.
 At baryon density $2\rho_0$ $(4\rho_0)$ the values of mass shift are found to be 
 108 (131), 104(128), 97(123) and 87(113) at 
  temperatures, T = 0, 50, 100 and 150 MeV.
  For the axial vector meson $B_{1}$
  at baryon density $\rho_0$ $(2\rho_0)$ the values of mass shift are found to be 
 267 (396), 251(381), 233(357) and 203(324) MeV at 
  temperatures, T = 0, 50, 100 and 150 MeV respectively. 
  At baryon density $4\rho_0$ the values of mass shift are found to be 
 492, 485, 467 and 434 MeV at 
  temperatures, T = 0, 50, 100 and 150 MeV respectively.
  The values of scattering lengths  for axial-vector mesons $D_1$ ($B_1$) at temperature T = 0
   and baryon densities, $\rho_B$ = $\rho_0$ and $4\rho_0$
  are observed to be 1.38 (6.02) and 0.62 (2.83) fm respectively.
For temperature T = 100 MeV and baryon densities, $\rho_B$ = $\rho_0$ and $4\rho_0$
the values of scattering lengths for  
$D^\star$ ($B^\star$) changes to  1.19 (5.25) and 0.58 (2.69) fm respectively.
  From the above discussions we observe a positive value of mass shift for
  scalar ($D_{0}$, $B_{0}$) and axial vector mesons ($D_{1}$, $B_{1}$) 
  in the nuclear medium. However the values of mass shift for
  vector mesons ($D^{\star}$, $B^{\star}$) are found to be negative.
  It means the masses of above scalar and axial-vector mesons in the nuclear
   medium may be large compared to the value in free space and this may lead to
   a decrease in the yield of these mesons in heavy-ion collisions.

\begin{figure}
 \resizebox{0.8\textwidth}{!}{%
 \includegraphics{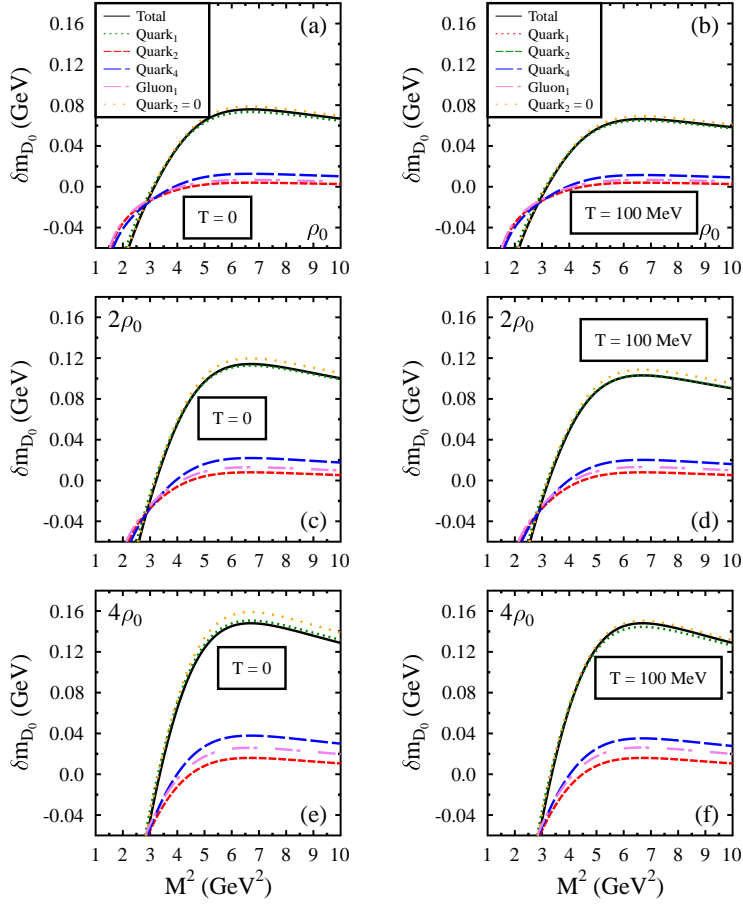}
}  
\caption{(Color online)
Figure shows the contribution of individual condensates to the mass shift of scalar mesons $D_{0}$ 
 as a function of the squared Borel mass parameter, $M^2$. The subplots (a), (c) and (e) show the
 results at temperature T = 0, whereas the
subplots (b), (d) and (f) are plotted for temperature T = 100 MeV.
We have shown the results at baryon densities $\rho_0$, $2\rho_0$ and $4\rho_0$.}
\label{scalarD0indfigg}
\end{figure}

\begin{figure}
  \resizebox{0.8\textwidth}{!}{%
 \includegraphics{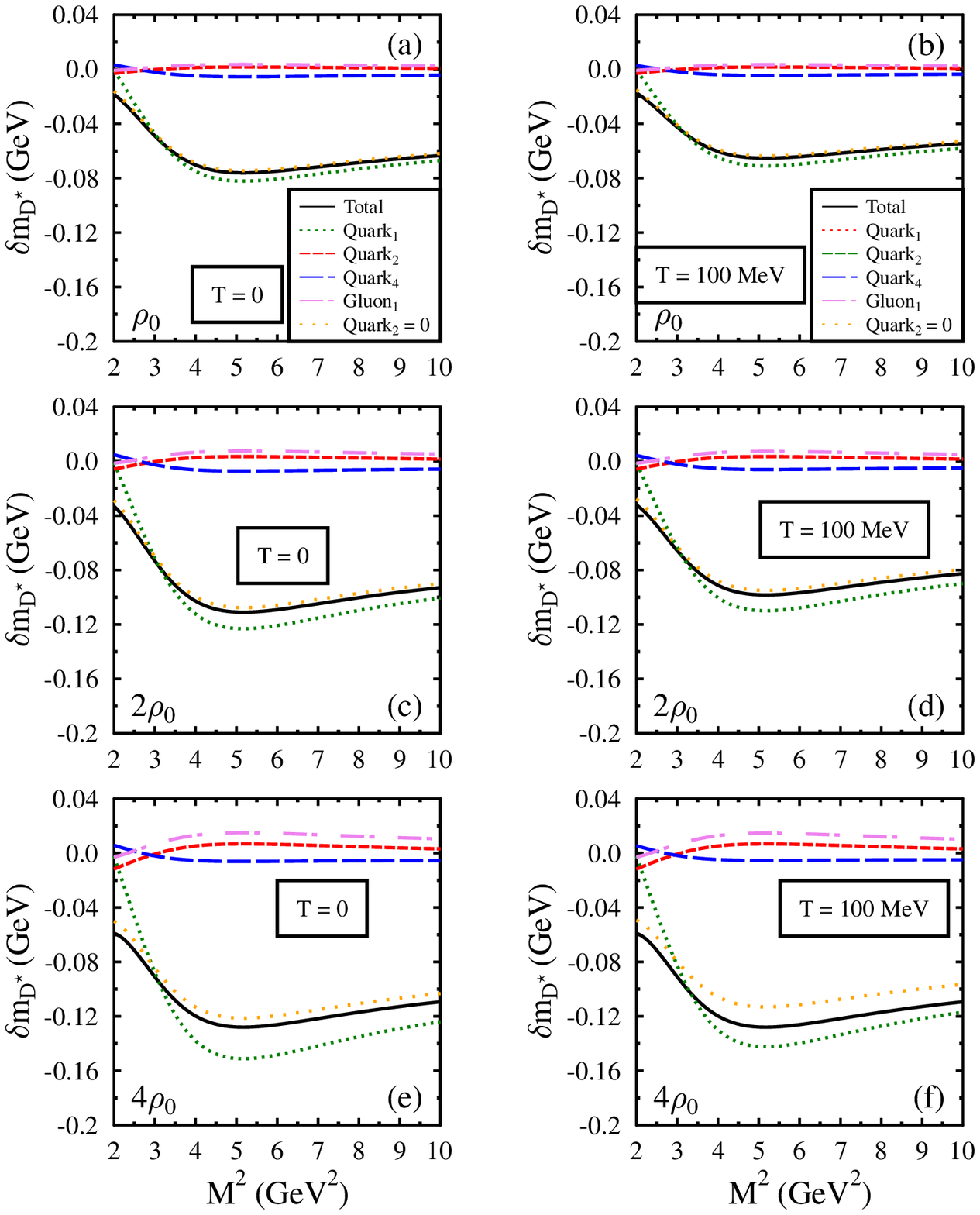}
}  
\caption{(Color online)
Figure shows the contribution of individual condensates to the mass shift of vector mesons $D^{\star}$ 
 as a function of the squared Borel mass parameter, $M^2$. The subplots (a), (c) and (e) show the
 results at temperature T = 0, whereas the
subplots (b), (d) and (f) are plotted for temperature T = 100 MeV.
We have shown the results at baryon densities $\rho_0$, $2\rho_0$ and $4\rho_0$.}
\label{vectorDstindfigg}
\end{figure}

\begin{figure}
    \resizebox{0.8\textwidth}{!}{%
 \includegraphics{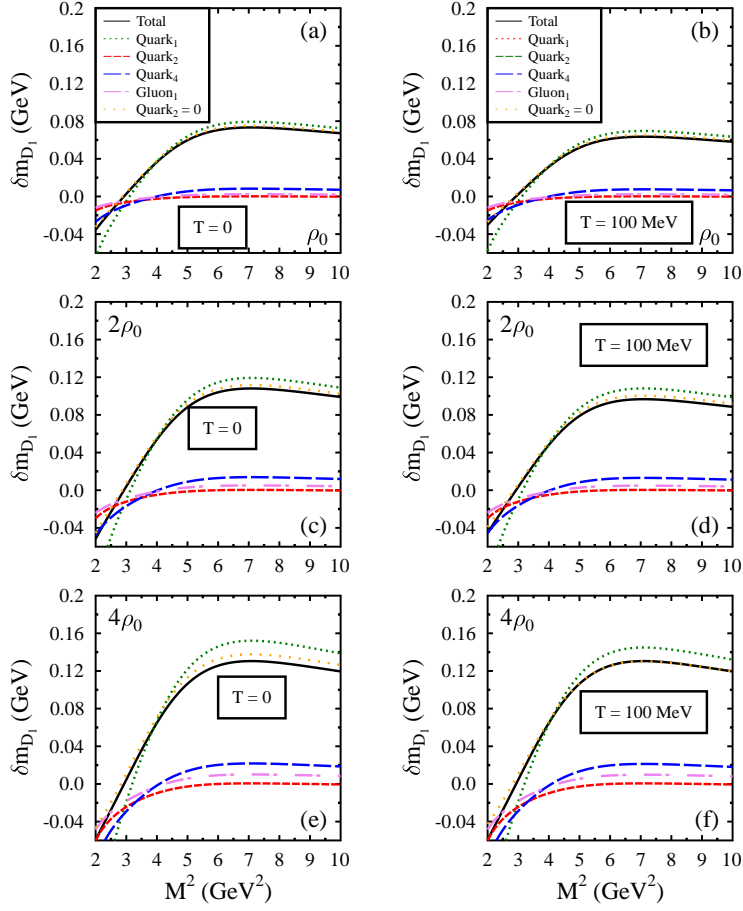}
}  
\caption{(Color online)
Figure shows the contribution of individual condensates to the mass shift of axial vector mesons $D_{1}$ 
 as a function of the squared Borel mass parameter, $M^2$. The subplots (a), (c) and (e) show the
 results at temperature T = 0, whereas the
subplots (b), (d) and (f) are plotted for temperature T = 100 MeV.
We have shown the results at baryon densities $\rho_0$, $2\rho_0$ and $4\rho_0$.}
\label{axialvectorD1indfigg}
\end{figure}

Now we shall discuss the effect of different individual condensates on the
in-medium modification of 
scalar ($D_{0}$, $B_{0}$) , vector  ($D^{\star}$, $B^{\star}$) and axial vector  ($D_{1}$, $B_{1}$) mesons.
In figures  (\ref{scalarD0indfigg}), (\ref{vectorDstindfigg}) and (\ref{axialvectorD1indfigg})
 we  compare the contributions of individual condensates to the mass shift
of scalar mesons, $D_0$, vector mesons, $D^{\star}$ and axial vector mesons, $D_{1}$ respectively.
 The subplots (a), (c) and (e) show the
results at temperature T = 0, whereas the
subplots (b), (d) and (f) are plotted for temperature T = 100 MeV.
We have shown the results at baryon densities $\rho_0$, $2\rho_0$ and $4\rho_0$. 
Note that in figures (\ref{scalarD0indfigg}) and in the subsequent figures of this paper
the word $``Total"$ is for the contribution of all condensates to the properties of 
mesons. Similarly, $``Quark_1"$, $``Quark_2"$, $``Quark_4"$ and $``Gluon_1"$
are denoting the contribution of $\langle\bar{q}q\rangle$, $\langle q^\dag i D_0q\rangle$,
 $\langle \bar{q} i D_0 i D_0q\rangle$
and $\langle\frac{\alpha_sGG}{\pi}\rangle$ respectively to the
in-medium properties of mesons.

\begin{figure}
   \resizebox{0.8\textwidth}{!}{%
 \includegraphics{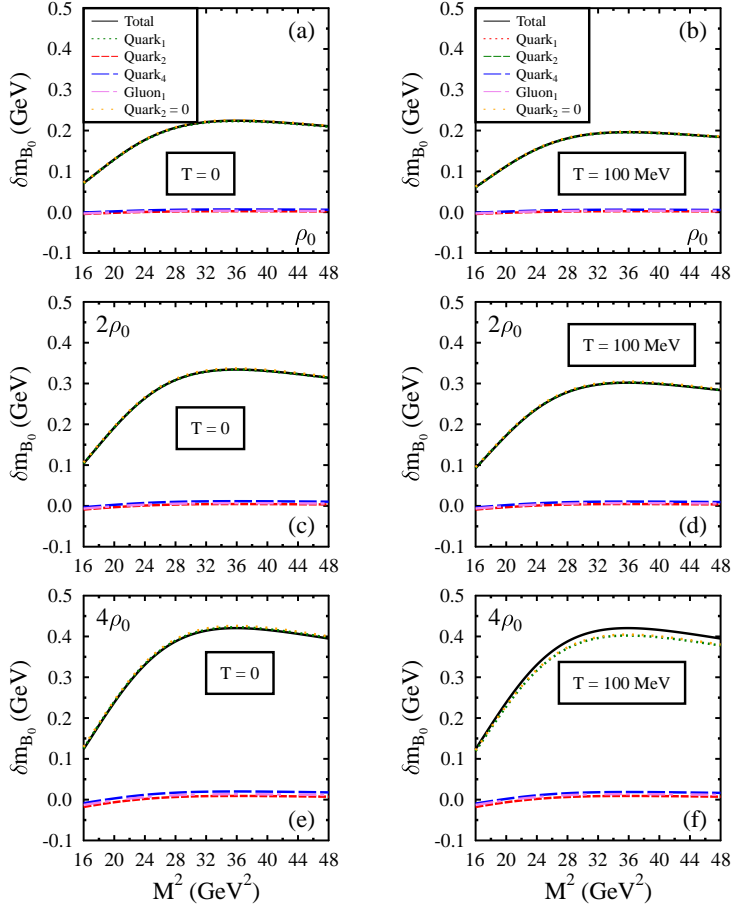}
} 
\caption{(Color online)
Figure shows the contribution of individual condensates to the mass shift of scalar mesons $B_{0}$ 
 as a function of the squared Borel mass parameter, $M^2$. The subplots (a), (c) and (e) show the
 results at temperature T = 0, whereas the
subplots (b), (d) and (f) are plotted for temperature T = 100 MeV.
We have shown the results at baryon densities $\rho_0$, $2\rho_0$ and $4\rho_0$. }
\label{scalarB0indfigg}
\end{figure}

\begin{figure}
   \resizebox{0.8\textwidth}{!}{%
 \includegraphics{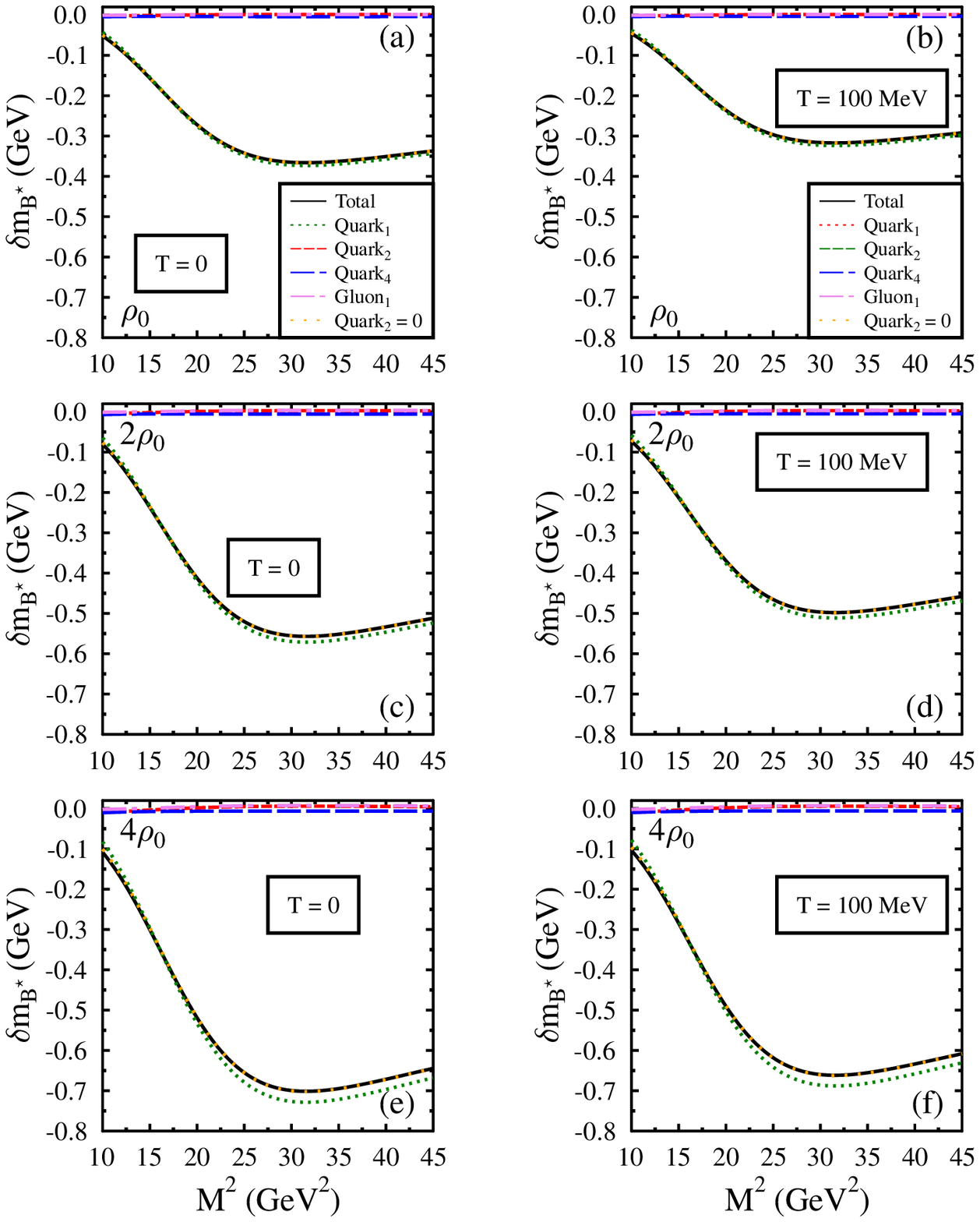}
}  
\caption{Figure shows the contribution of individual condensates to the mass shift of vector mesons $B_{\star}$ 
 as a function of the squared Borel mass parameter, $M^2$. The subplots (a), (c) and (e) show the
 results at temperature T = 0, whereas the
subplots (b), (d) and (f) are plotted for temperature T = 100 MeV.
We have shown the results at baryon densities $\rho_0$, $2\rho_0$ and $4\rho_0$.}
\label{vectorBstindfigg}
\end{figure}

\begin{figure}
 \resizebox{0.8\textwidth}{!}{%
 \includegraphics{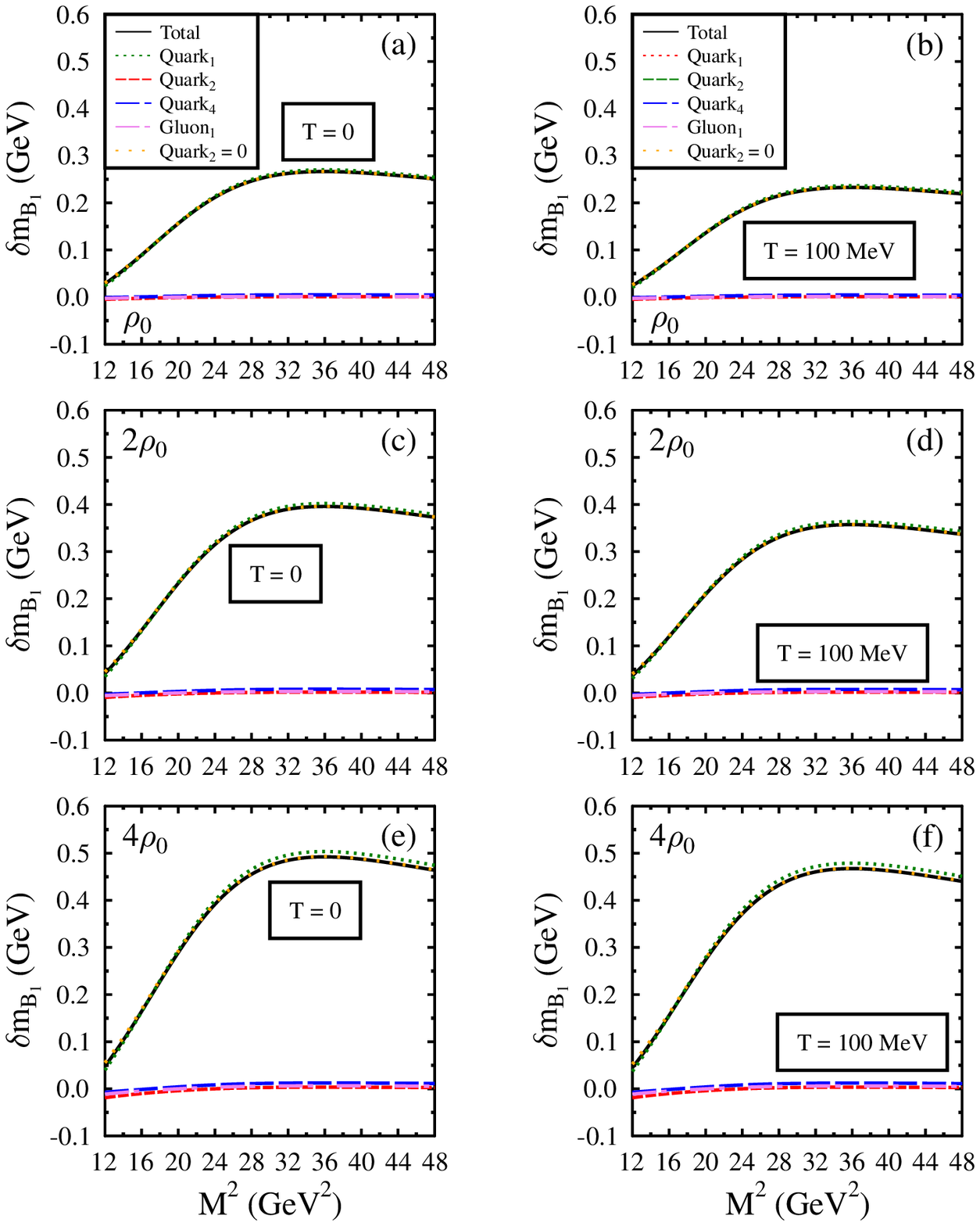}
}  
\caption{(Color online)
Figure shows the contribution of individual condensates to the mass shift of axial vector mesons $B_{1}$ 
 as a function of the squared Borel mass parameter, $M^2$. The subplots (a), (c) and (e) show the
 results at temperature T = 0, whereas the
subplots (b), (d) and (f) are plotted for temperature T = 100 MeV.
We have shown the results at baryon densities $\rho_0$, $2\rho_0$ and $4\rho_0$.}
\label{axialvectorB1indfigg}
\end{figure}

 We observe that at temperature T =  0 
 if we consider the contribution of scalar quark condensates only then the values
 of mass shift for the scalar $D_0$ mesons are found to be $73.30$ and $150.68$  MeV 
 at nuclear matter density $\rho_B$ = $\rho_0$ and $4\rho_0$ respectively.
 When we consider the individual contributions of 
  $\langle q^\dag i D_0q\rangle$, $\langle \bar{q} i D_0 i D_0q\rangle$
and $\langle\frac{\alpha_sGG}{\pi}\rangle$
 condensates then the values of mass-shift at $\rho_0$ ($4\rho_0$) and T = 0
  are observed to be  4.01(16.07), 12.70(37.87) and 6.68(25.96) MeV
 respectively.
 From above discussion we observe that the maximum contribution to the in-medium modification
 of scalar $D_0$ mesons is from light quark condensate  $\langle \bar{q}q\rangle$.
 Note that leaving the quark condensate, $\langle q^\dag i D_0q\rangle$, the
 all other condensates have been evaluated within chiral SU(3) model in the present
 investigation. So in figure  (\ref{scalarD0indfigg}) we also show the
 variation of the 
 mass shift of scalar $D_0$ mesons as a function of squared Borel mass parameter
 when we neglect the contribution of $\langle q^\dag i D_0q\rangle$  condensate.
 We observe that if we neglect the contribution of $\langle q^\dag i D_0q\rangle$
 condensate then the values of mass shift are found to be
 78.70 and  158.99 MeV at densities $\rho_0$ and $4\rho_0$ respectively.
 Note that considering the contribution of all condensates, at
 temperature T = 0, the values of mass shift were 75.88 and 148.05 MeV at 
 densities $\rho_0$ and $4\rho_0$ respectively.
 So we observe that if we neglect the  condensate
  $\langle q^\dag i D_0q\rangle$  then there is a percentage 
 change of 4 $\%$ and 7 $\%$ in the mass shift of $D_0$ mesons at densities $\rho_0$ and $4\rho_0$ respectively.
 From figure (\ref{vectorDstindfigg}), we 
 observe that for vector mesons, $D^{*}$ at temperature T = 0 and
  baryon density $\rho_B$ = $\rho_0$ ($4\rho_0$), the values of mass shift 
  due to condensates $\langle\bar{q}q\rangle$, $\langle q^\dag i D_0q\rangle$,
$\langle\bar{q}g_s\sigma Gq\rangle$, $\langle \bar{q} i D_0 i D_0q\rangle$
and $\langle\frac{\alpha_sGG}{\pi}\rangle$ are
observed to be -82.15(-151.15), 1.689(6.749), 
18.95(57.16), -5.49(-6.16) and 3.54 (14.93)  MeV respectively.
For temperature T = 100 MeV the above values of mass shift at 
baryon density $\rho_B$ = $\rho_0$ ($4\rho_0$) changes to 
 -73.01 (-142.35), 1.689 (6.75), 17.85 (56.34), -4.53 (-5.43) and
 3.43 (14.64) MeV respectively.
For the axial vector mesons, $D_1$  the values of mass shift due to individual condensates 
$\langle\bar{q}q\rangle$, $\langle q^\dag i D_0q\rangle$,
$\langle\bar{q}g_s\sigma Gq\rangle$, $\langle \bar{q} i D_0 i D_0q\rangle$
and $\langle\frac{\alpha_sGG}{\pi}\rangle$ are
observed to be 79.31(152), 0.1516(0.607), -8.77(-22.63), 8.27(21.76) and 2.34 (10.08)  MeV respectively.
For temperature T = 100 MeV the above values of mass shift at 
baryon density $\rho_B$ = $\rho_0$ ($4\rho_0$) changes to 
 70 (145), 0.151 (0.607), -7.99(-22.04), 7.58 (21.26) and
 2.23 (9.81) respectively.
 In figures  (\ref{scalarB0indfigg}), (\ref{vectorBstindfigg}) and (\ref{axialvectorB1indfigg})
 we have shown the contributions of individual condensates to the mass shift
 of scalar mesons, $B_0$, vector mesons, $B^{\star}$ and axial vector mesons, $B_{1}$ respectively.
 The subplots (a), (c) and (e) show the
results at temperature T = 0, whereas the
subplots (b), (d) and (f) are plotted for temperature T = 100 MeV.
We have shown the results at baryon densities $\rho_0$, $2\rho_0$ and $4\rho_0$. 
 In table  (\ref{scalarB0indtab}), (\ref{vectorBstindtab}) and (\ref{axialvectorB1indtab})
 we have tabulated the values of mass shift for 
  scalar mesons, $B_0$, vector mesons, $B^{\star}$ and axial vector mesons, $B_{1}$ respectively.
  The values of mass shift have been given at baryon densities $\rho_0$ and $4\rho_0$	
  and temperatures T = 0 and 100 MeV.
  \begin{table}
\begin{tabular}{|l|l|l|l|l|}
\hline
 & \multicolumn{2}{c|}{T = 0} & \multicolumn{2}{c|}{T = 100 MeV} \\ 
\hline
 & $\rho_B = \rho_0$ & $\rho_B = 4\rho_0$ & $\rho_B = \rho_0$ & $\rho_B = 4\rho_0$ \\ 
\hline
Total & 224 & 420 & 196 & 399 \\ 
\hline
$\langle\bar{q}q\rangle$ & 223 & 422 & 195 & 402 \\ 
\hline
$\langle q^\dag i D_0q\rangle$ & 2.26 & 9.04 & 2.26 & 9.04  \\ 
\hline
$\langle \bar{q} i D_0 i D_0q\rangle$ &6.60 & 20.04 & 5.99 & 18.69 \\ 
\hline
$\langle\frac{\alpha_sGG}{\pi}\rangle$ & 3.49 & 13.35 & 3.58 & 13.57 \\ 
\hline
$\langle q^\dag i D_0q\rangle$ = 0 & 226 & 426 & 197 & 405 \\ 
\hline
\end{tabular}
\caption{
The above table gives the values of mass shift of scalar mesons $B_{0 }$ (in MeV units) 
 due to individual condensates at temperatures T = 0 and 100 MeV.
 For each value of temperature the values are tabulated 
 for baryon densities $\rho_B = \rho_0$ and $4\rho_0$.}
 \label{scalarB0indtab}
\end{table}
 \begin{table}
\begin{tabular}{|l|l|l|l|l|}
\hline
 & \multicolumn{2}{c|}{T = 0} & \multicolumn{2}{c|}{T = 100 MeV} \\ 
\hline
 & $\rho_B = \rho_0$ & $\rho_B = 4\rho_0$ & $\rho_B = \rho_0$ & $\rho_B = 4\rho_0$ \\ 
\hline
Total & -367 & -705.69 & -318.08 & -665.88 \\ 
\hline
$\langle\bar{q}q\rangle$ & -373.11 & -728.58 & -323.61 & -687.98 \\ 
\hline
$\langle q^\dag i D_0q\rangle$ & 1.477 & 5.9057 & 1.477 & 5.9057 \\ 
\hline
$\langle\bar{q}g_s\sigma Gq\rangle$ & 12.43 & 36.77 & 11.67 & 36.18 \\ 
\hline
$\langle \bar{q} i D_0 i D_0q\rangle$ & -3.98 & -6.42 & -3.40 & -5.98 \\ 
\hline
$\langle\frac{\alpha_sGG}{\pi}\rangle$ & 1.64 & 7.56& 1.51 & 7.21 \\ 
\hline
$\langle q^\dag i D_0q\rangle$ = 0 & -366.34 & -702.26 & -317.28 & -662.48 \\ 
\hline
\end{tabular}
\caption{
The above table gives the values of mass shift of vector mesons $B^{\star}$ (in MeV units) 
 due to individual condensates at temperatures T = 0 and 100 MeV.
 For each value of temperature the values are tabulated 
 for baryon densities $\rho_B = \rho_0$ and $4\rho_0$.}
\label{vectorBstindtab}
\end{table}

\begin{table}
\begin{tabular}{|l|l|l|l|l|}
\hline
 & \multicolumn{2}{c|}{T = 0} & \multicolumn{2}{c|}{T = 100 MeV} \\ 
\hline
 & $\rho_B = \rho_0$ & $\rho_B = 4\rho_0$ & $\rho_B = \rho_0$ & $\rho_B = 4\rho_0$ \\ 
\hline
Total & 266.51 & 492.22 & 232.75 &467.29 \\ 
\hline
$\langle\bar{q}q\rangle$ & 270.11 & 503.57 & 236.35 &478.86 \\ 
\hline
$\langle q^\dag i D_0q\rangle$ &.88 & 3.53 & 0.88 & 3.53 \\ 
\hline
$\langle\bar{q}g_s\sigma Gq\rangle$ & -6.34 & -16.79 & -5.83 & -16.40 \\ 
\hline
$\langle \bar{q} i D_0 i D_0q\rangle$ & 4.79 & 12.53 & 4.39 & 12.23 \\ 
\hline
$\langle\frac{\alpha_sGG}{\pi}\rangle$ & 1.41 & 6.39 & 1.31 & 6.12 \\ 
\hline
$\langle q^\dag i D_0q\rangle$ = 0 & 266.75 & 493.11 & 232.99 & 468.20 \\ 
\hline
\end{tabular}
\caption{(Color online)
The above table gives the values of mass shift of axial vector mesons $B_{1}$ (in MeV units) 
 due to individual condensates at temperatures T = 0 and 100 MeV.
 For each value of temperature the values are tabulated 
 for baryon densities $\rho_B = \rho_0$ and $ 4\rho_0$.}
 \label{axialvectorB1indtab}
\end{table}

In ref. \cite{wang1} the properties of scalar mesons $D_{0}$ and $B_0$ 
 had been studied using QCD sum rules
and the observed values of mass shift at nuclear saturation 
density were $69$ and $217$ MeV respectively.
The properties of $D_0$ mesons had also been studied in ref. \cite{tolos1}
using coupled channel approach. There an extra widening
 from the already large width of the resonance in free space
was observed for the $D_0$ meson. 
The properties of vector mesons ($D^*$ and $B^*$)  and axial-vector ($D_1$ and  $B_1$)
had been studied using QCD sum rules in ref. \cite{wang2}.
For vector mesons $D^*$ and $B^*$ the values of mass shift  
were $-71$ and $-380$ MeV respectively whereas for axial vector mesons
$D_1$ and  $B_1$ these values changes to $72$ and $264$ MeV respectively. 
The observed positive values of mass shift for $D_{0}$ and $D_{1}$ mesons
in our present investigation and also in earlier investigations 
indicate that these mesons feels repulsive interactions in the
nuclear medium and their in-medium mass increases as function of baryonic density. This means the chances of 
decay of higher charmonium states to these heavy charmed mesons pairs
are suppressed and hence these mesons may not cause a decrease in 
the production of $J/\psi$ mesons in heavy-ion collisions.
However for the vector mesons $D^*$
we observe the negative values of mass shift and hence we conclude that they feel
attractive interactions in the nuclear medium. 
The decrease in the mass of vector mesons $D^*$ may cause the 
decay of higher charmonium states to $D^*\bar{D}^*$ pairs
and hence it may be a cause of  $J/\psi$ suppression in
heavy-ion collision experiments.

\section{Summary}
In the present paper we investigated the mass modifications of 
scalar mesons $\left( D_{0}, B_{0}\right)$,
 vector mesons $\left( D^{\ast}, B^{\ast}\right)$ 
and axial vector mesons $\left( D_{1}, B_{1}\right)$ at
 finite density and temperature of the nuclear medium.
We used QCD sum rules along with chiral SU(3) model to investigate the
properties of above mentioned mesons.
Using chiral SU(3) model we found the values of quark and gluon condensates
at finite density and temperature of the 
medium which were further used within QCD sum rules
to find the in-medium masses of scalar, vector and axial vector mesons.
We observed a positive value of  mass shift for the scalar mesons ($D_0$ and $B_0$)
 and axial vector mesons ($D_1$ and $B_1$)
 i.e. their in-medium mass found to be more as compared to the 
vacuum mass. For a constant value of temperature the values of mass shift for these
scalar and axial-vector mesons are found to 
increase as we increase the density of the nuclear medium.
On the other hand for a constant value of density the temperature of the 
medium causes a decrease in the values of mass shift of  scalar and axial-vector mesons.
 The vector mesons  $D^\star$ and $B^\star$ are found to have negative values of mass shift.
It means their in-medium masses are small as compared to vacuum masses.
For a constant value of density, as a function of temperature, the magnitude of
  mass shift of $D^\star$ and $B^\star$
mesons decreases. However, for a constant value of temperature, as a
function of density the 
magnitude of mass shift  of vector mesons $D^\star$ and $B^\star$ 
are found to increase.
We have also investigated the effects of individual terms on the mass shift of  mesons.
It was found that the scalar quark condensates, $\bar{q}q$ has maximum contribution to 
the in-medium modification of scalar, vector and axial vector mesons.
The effects of density and temperature of the medium on the scattering lengths
of scalar, vector and axial vector mesons were also investigated.
We observed the positive value of scattering lengths for scalar and axial vector mesons
whereas for the vector mesons a negative value of
scattering lengths
were observed. Also it was found that the
magnitude of the scattering lengths of scalar, vector and axial-vector 
mesons decreases as we move from low to high value of
density or temperature of the nuclear medium.
The present investigation of medium modification of scalar, vector and 
axial vector mesons will be helpful for understanding their production
rate and also the phenomenon of $J/\psi$ suppression in the 
Compressed Baryonic matter experiment at FAIR, GSI.

\acknowledgements 

Author is thankful to Amruta Mishra for fruitful discussions 
on chiral SU(3) model and QCD sum rules. The financial support from
the Department of Science and Technology (DST), India for research project under
 fast track scheme for young scientists (SR/FTP/PS-209/2012)
is gratefully acknowledged.

\end{document}